\documentclass[a4paper,11pt]{article}
\pdfoutput=1 

\usepackage{jheppub} 

\usepackage{latexsym,amssymb,amstext,amsmath}
\usepackage{hyperref}
\usepackage{graphicx} 
\usepackage{subcaption}
\usepackage{amsthm}
\usepackage{esint}
\usepackage{float}

\def\be {\begin{equation}}
\def\ee {\end{equation}}
\def\bea {\begin{eqnarray}}
\def\eea {\end{eqnarray}}
\def\nn {\nonumber}

\def\beq{\begin{equation}}
\def\eeq{\end{equation}}
\def\beqa{\begin{eqnarray}}
\def\eeqa{\end{eqnarray}}

\newcommand{\ft}[2]{{\textstyle\frac{#1}{#2}}}

\begin{document}
%
\begin{titlepage}

\begin{center}
 {\LARGE\bfseries 
 Observations on holographic aspects of \\  four-dimensional asymptotically flat  \\ 
 \vskip 2mm ${\cal N}=2$ black holes}
 \\[10mm]

\textbf{Pedro Aniceto, Gabriel Lopes Cardoso and Suresh Nampuri }

\vskip 6mm
{\em  Center for Mathematical Analysis, Geometry and Dynamical Systems,\\
  Department of Mathematics, 
  Instituto Superior T\'ecnico,\\ Universidade de Lisboa,
  Av. Rovisco Pais, 1049-001 Lisboa, Portugal}\\
\vskip 3mm

{\tt 
pedro.aniceto@tecnico.ulisboa.pt,
gabriel.lopes.cardoso@tecnico.ulisboa.pt}, \\
{\tt  nampuri@gmail.com}
\end{center}

\vskip .2in
\begin{center} {\bf ABSTRACT } \end{center}
\begin{quotation}

\noindent 
In this note, we explore  holographic attributes of
four-dimensional near-extremal Reissner-Nordstrom black hole solutions in ungauged ${\cal N}=2$  supergravity theories at the two-derivative level by
recasting them  
as a specific
first-order deformation in solution space, associated with an infinitesimal Harrison transformation,
of black holes in an $AdS_2$ space-time.  Specifically, we use this link to exhibit how bulk properties, such as mass and entropy, 
of four-dimensional near-extremal
black holes 
are holographically encoded in the one-dimensional boundary theory
dual to gravity in an infinitesimally deformed $AdS_2$ space-time. 
We do so for the case of four-dimensional near-extremal black holes that arise as deformations in solution space of BPS black holes 
by changing the non-extremality parameter. For these near-extremal black holes, we 
further show that the 
nAdS$_2$ attractor mechanism
can be recast as a specific deformation of 
the BPS flow equations in four dimensions. Additionally, we also discuss
time-dependent perturbations of the four-dimensional 
near-extremal 
Reissner-Nordstrom solutions from a two-dimensional point of view.

\vskip 3mm
\noindent

\end{quotation}
\vfill
\today
\end{titlepage}

\tableofcontents

\section{Introduction}

In this note we discuss aspects of the holographic correspondence for
asymptotically flat four-dimensional 
near-extremal black hole solutions 
that arise in four-dimensional 
ungauged ${\cal N}=2$  supergravity theories
at the two-derivative level. We do so by first reducing these theories down to two dimensions on a two-sphere, and subsequently viewing 
the four-dimensional solutions as solutions to the equations of motion of the theory in two dimensions.

It is well known that the above class of two-derivative four-dimensional gravitational theories admits two distinct types of vacua and hence, 
correspondingly, two types of four-dimensional static black hole solutions based on these spatial asymptotia or vacua \cite{Gibbons:1982fy,Gibbons:1985bz}.  
The first corresponds to an asymptotic $AdS_2$ $\times S^2$ 
space-time and includes two-dimensional black holes in the $AdS_2$ geometry \cite{Spradlin:1999bn}. The AdS/CFT correspondence posits a description of these backgrounds in a holographically dual one-dimensional CFT.  
The second corresponds to four-dimensional asymptotic Minkowski space-time
and includes static extremal and non-extremal black hole solutions which are supported by the scalar fields in the theory. 
The radial flow of the scalar fields from the vacuum to the black hole horizon is governed by an effective black hole potential generated by the 
$U(1)$ gauge fields in the theory.

In the case of extremal black holes, whose near horizon geometry is the vacuum $AdS_2 \times S^2$ background and hence can be seen as solitonic solutions interpolating between the two vacua \cite{Gibbons:1982fy,Gibbons:1985bz}, the scalar fields are fixed in terms of the charges at the black hole horizon, independent of their asymptotic values,  i.e., the extremal black hole near-horizon geometry acts as an attractor for the scalar field flows 
\cite{Ferrara:1995ih,Ferrara:1996dd,Ferrara:1997tw,Goldstein:2005hq}. This attractor mechanism serves as a local test for the presence of an extremal black horizon. This corresponds to the black hole potential being extremized at the attractor point. Therefore, if the asymptotic values of the scalar fields for an extremal black hole with given charges are chosen to lie at their attractor values,
the extremal black solution is supported by constant physical scalar fields and the resulting solution is referred to as double-extreme, with a geometry described by a 
four-dimensional extremal Reissner-Nordstrom black hole solution. The near-horizon geometry containing an $AdS_2$ factor, wherein all geometrical scales as well as physical fields are determined purely by the charges independently of the asymptotic vacuum, implies that the extremal black hole entropy 
must be fully captured by the near-horizon $AdS_2$ factor, and hence admit a holographic description in terms of the dual one-dimensional CFT. However, non-extremal static four-dimensional black holes have a Rindler  near-horizon geometry and cannot be seen  as interpolating solutions between vacua, precluding a clear template for a holographic description.  These solutions are characterized by the extremal black hole parameters, such as charge, in addition to a non-extremality parameter $\mu$ which encodes the temperature of the black hole and can be viewed as one of the axes in the 
asymptotically flat black hole solution space of the theory. Moving along this axes from the origin corresponds to an on-shell deformation starting from an extremal black hole and sliding through non-extremal black holes at different temperatures and different $\mu$-dependent horizon location and entropy, but with the same charges. 

In this note, we take preliminary steps in exploring the holographic description of the simplest types of non-extremal black holes, namely near-extremal solutions with $\mu \ll 1$ and supported by constant physical scalar fields. These have non-extremal Reissner-Nordstrom metrics which 
we view as on-shell deformations to first order in $\mu$ from BPS double-extreme solutions. Near-extremal black hole solutions are
in effect, the 'closest' non-extremal black solutions in solution space to the double-extreme black holes. At this order, the near-extremal Reissner-Nordstrom black hole solution has a thermodynamic temperature that is proportional to $\mu$, its entropy is shifted by a term proportional to $\mu$ relative 
 to the entropy of the extremal solution, while its mass is shifted by a term proportional to  $\mu^2$ relative to the mass
 of the extremal solution. The equations of motion of the theory, and hence the solution space, are invariant under a one-parameter family of  transformations \cite{Goldstein:2014gta} called Harrison transformations, which map the above near-extremal Reissner-Nordstrom solution to a 
 two-dimensional black hole with an asymptotic $AdS_2$ $\times S^2$ geometry, namely the first type of aforementioned solutions. The horizon features of the four-dimensional black hole and its  two-dimensional Harrison image are identical. Denoting the transformation parameter by $\varepsilon$, we can hence model the four-dimensional near-extremal Reissner-Nordstrom black hole of interest as a specific infinitesimal Harrison transformation of a 
 two-dimensional black hole in an $AdS_2$ space-time. The infinitesimal transformation acts as a deformation in solution space, implemented to first order in $\varepsilon$ and $\mu$.

Using this link, we ascribe mass to the deformed two-dimensional solution, and we show that this definition is consistent with the 
first law of black hole mechanics, which relates the change of the entropy of the deformed two-dimensional solution to the change of its mass.

We then exploit the link between four-dimensional and two-dimensional solutions provided by the Harrison transformation 
as a first step to exploring the holographic attributes of
four-dimensional near-extremal Reissner-Nordstrom black hole solutions. 
By virtue of the AdS/CFT correspondence, we expect the two-dimensional black hole parameters and their perturbations to be encoded in terms of corresponding operators and their perturbations in the holographically dual quantum mechanics. Consequently, 
we exhibit how bulk properties of the four-dimensional near-extremal 
Reissner-Nordstrom black hole, such as mass and entropy, are encoded in terms of 
the expectation value of the holographic stress tensor of the one-dimensional boundary theory and 
of the operator dual to the two-dimensional dilaton, respectively.

We continue our exploration of this holographic setup by considering
time-dependent perturbations of the four-dimensional 
near-extremal 
Reissner-Nordstrom solutions from a two-dimensional point of view. To this end, we use the two-dimensional equations of motion to analyze first-order
perturbations around two-dimensional constant curvature solutions that are supported by constant scalar fields as well as electric-magnetic charges.
Our findings for the allowed first-order perturbations are
the counterpart of those obtained previously in a different context, 
namely
five-dimensional pure Einstein gravity with cosmological constant
reduced to two dimensions \cite{Castro:2018ffi}, 
and three-dimensional topologically massive gravity reduced to two dimensions \cite{Castro:2019vog}.
When restricting to the subset of perturbations that are induced solely by turning on a perturbation of the two-dimensional dilaton,
the renormalized on-shell action describing the dynamics of these perturbations takes the form of the DFF action  \cite{deAlfaro:1976vlx}.
This is again in accord with the findings in \cite{Castro:2018ffi,Castro:2019vog}.

While in the above we focussed on near-extremal black hole solutions in four dimensions that are supported
by physical scalar fields that are constant, one may also consider the more general situation where the scalar fields supporting
the near-extremal black hole solution are not any longer constant. This may be achieved by changing the asymptotic values
of the scalar fields away from the double-extreme values mentioned earlier. The resulting flow equations can be viewed as deformations of the flows in the extremal case. In the case where these near-extremal solutions are on-shell deformations of extremal black holes, 
a suitable framework for describing the resulting scalar flows 
is the nAdS$_2$ attractor mechanism for near-extremal black holes \cite{Larsen:2018iou}. It
posits that to first order in the non-extremality parameter $\mu$, the near-horizon behaviour of the 
metric and of the scalar fields of the non-extremal black hole solution is governed by the near-horizon behaviour of the 
associated extremal
black hole solution. In the case when the near-extremal black hole solution is an on-shell deformation of a BPS black hole solution, 
we show how the nAdS$_2$ attractor mechanism for near-extremal black holes \cite{Larsen:2018iou} is encoded
in deformed BPS flow equations. We do this by
using the formalism of Hamilton's principal function in the two-dimensional theory, to rewrite the standard flow equations for four-dimensional BPS solutions \cite{Ferrara:1995ih,Ferrara:1997tw} 
in
ungauged ${\cal N}=2$  supergravity theories in terms of a real quantity $W$. Subsequently, by performing a judicious 
deformation of these flow equations, we show how the nAdS$_2$ attractor mechanism for near-extremal black holes 
arises at lowest order in the deformation parameter $\mu$ and infinitesimally away from the horizon.

As mentioned above, in this note we focus on near-extremal Reissner-Nordstrom solutions that we regard as 
first-order deformations of four-dimensional BPS backgrounds in ungauged ${\cal N}=2$  supergravity theories at the two-derivative
level. We then explore some of their holographic attributes by viewing them 
as a specific first-order perturbation of a two-dimensional black hole in an $AdS_2$ space-time. We refer to \cite{Castro:2021wzn}
for further work on the 
holographic properties of near-extremal solutions in ungauged ${\cal N}=2$  supergravity theories in four dimensions.
 It would be interesting to extend the analysis given here by including $R^2$-corrections based on $F$-terms. While their effect
on BPS solutions in the near-horizon region is well studied \cite{LopesCardoso:2000qm,Cardoso:2006xz}, this is not the case for near-extremal black hole solutions.
It would also be interesting to further extend the analysis by introducing further deformations, for instance deformations that break spherical
symmetry. While in this 
note we focussed on near-extremal solutions in ungauged ${\cal N}=2$  supergravity theories in four dimensions, work on 
holographic properties of near-extremal solutions in gauged ${\cal N}=2$  supergravity theories in four dimensions has 
recently appeared in \cite{Castro:2021wzn}.

We expect the holographic setup outlined in this note to indicate preliminary steps towards 
improving the nAdS$_2$/nCFT$_1$ dictionary
 \cite{Almheiri:2014cka,Maldacena:2016upp,Sarosi:2017ykf,Brown:2018bms,Cvetic:2016eiv,Castro:2018ffi,Castro:2019vog,Castro:2021wzn}
 for four-dimensional asymptotically flat black holes.

\section{Dimensionally reduced two-dimensional bulk Lagrangian } \label{sec_dimensionally_reduced_theory}

In this section, we review the dimensional reduction of $\mathcal{N} = 2$ supergravity in four dimensions
at the two-derivative level on a two-sphere down to two dimensions.  This dimensional reduction has been discussed in the literature,
both in the context of Poincar\'e supergravity theories (see, for instance, \cite{Larsen:2018iou}) as well as in the context of the superconformal approach to
$\mathcal{N} = 2$ supergravity 
\cite{Cardoso:2006xz}.  Here, we follow \cite{Cardoso:2006xz}.
The resulting two-dimensional action
is a generalization of the two-dimensional action that results when dimensionally reducing the four-dimensional Einstein-Maxwell theory on
a two-sphere \cite{Sarosi:2017ykf,Brown:2018bms}.

We consider the Wilsonian Lagrangian describing the coupling of Abelian vector multiplets to $\mathcal{N} = 2$ supergravity in four dimensions
at the two-derivative level.  The associated Lagrangian can be constructed using the superconformal multiplet calculus \cite{deWit:1984rvr}.
This requires adding two compensating multiplets, whose role is to ensure that upon gauge fixing, the resulting theory is a Poincar\'e supergravity
theory. One of these compensating multiplets is a vector multiplet, while the other compensating multiplet can be taken to be a hyper multiplet.
Below we will display the resulting bosonic part of this Lagrangian in Poincar\'e gauge.

As is well-known \cite{deWit:1984wbb}, this Lagrangian is encoded in a 
holomorphic function $F(X)$, which 
is
homogeneous of second degree under scalings by $\lambda \in \mathbb{C} \backslash \{0\}$, 
\bea
F(\lambda X) = \lambda^2 F(X) \;.
\label{prep}
\eea
Here, the $X^ I$ ($I = 0, \dots, n$) denote
complex scalar fields that reside in the Abelian vector multiplets (which include the compensating vector multiplet).
We denote $F_I = \partial F(X)/\partial X^I$, $F_{IJ} = \partial^2 F(X) / \partial  X^I \partial X^J$.
The Poincar\'e gauge condition is given by
\bea
e^{- {\cal K} (X)} \equiv i \left( {\bar X}^I F_I - {\bar F}_I X^I \right) = \frac{1}{G_4} \;,
\label{Pgc}
\eea
where $G_4$ denotes Newton's constant in four dimensions.  In Poincar\'e gauge, the bosonic part of the Lagrangian 
 is given by
\cite{LopesCardoso:2000qm}
\begin{eqnarray}
  \label{eq:efflag}
  8\pi\, e^{-1} \, L &=&   - \frac{1}{2 G_4}  R + \left[ 
   i {\cal D}^{\mu} F_I \, {\cal D}_{\mu} \bar X^I   
   \right.
    \nonumber\\
  && \qquad \qquad 
  +\ft14 i F_{IJ} (F^{-I}_{ab} -\ft 14 \bar X^I 
  T_{ab}^-  ) (F^{-Jab} -\ft14 \bar X^J 
  T^{ab -} )  \nonumber\\
  &&
   \qquad \qquad 
   \left.
  -\ft18 i F_I(F^{+I}_{ab} -\ft14  X^I 
  T_{ab}^+) \, T^{ab+} 
    -\ft1{32} i F \, \bar{{\hat A}} + {\rm h.c.} 
    \right] \;,
\end{eqnarray}
where
\bea
 {\cal D}_{\mu} X^I &=& \partial_{\mu} X^I + i A_{\mu} X^I \;, \nonumber\\
 {\cal D}_{\mu} F_I &=& \partial_{\mu} F_I + i A_{\mu} F_I\;, \nonumber\\
 \hat A &=& (T_{ab}^-)^2 \;,
 \eea
 with \eqref{eq:efflag} subjected to the condition \eqref{Pgc}.
 The 
$U(1)$ connection $A_{\mu}$,  when using its equation of motion, becomes expressed as 
\bea
A_{\mu} = \tfrac12 e^{{\cal K}(X)} \left( F_I \partial_{\mu} {\bar X}^I - {\bar X}^I \partial_{\mu} F_I + c.c. \right) \;,
\label{u1conn}
\eea
which is subjected to the condition \eqref{Pgc}.

We will consider the dimensional reduction of the above theory on a two-sphere $S^2$, 
\bea
ds_4^2 = ds_2^2 + v_2  \, d \Omega_2^2 \;,
\eea
where $ d \Omega_2^2$ denotes the line element of the two-sphere $S^2$, and 
where $v_2 $ depends on the coordinates of the two-dimensional line element $ds_2^2$.
We denote the local coordinates on $S^2$ by $(\theta, \varphi)$, and local
coordinates in two space-time dimensions by $(r,t)$.  In a  spherically symmetric configuration, 
$T_{ab}^-$ is expressed in terms of a single complex scalar field $w$ \cite{Cardoso:2006xz}, 
\begin{equation}
  \label{eq:T-w}
  T_{\underline{r}\underline{t}}^{-}  =- i
  T_{\underline{\theta}\underline{\varphi}}^{-} =
  w \;,
 \end{equation}
where ${\underline{r}\underline{t}}, {\underline{\theta}\underline{\varphi}}$ denote Lorentz indices.
Then, 
\bea
\hat A = - 4 w^2 \;.
\eea

The four-dimensional backgrounds we will consider are supported by complex scalar fields $X^I$ and $w$, as well as
by electric fields $e^I$ and magnetic charges $p^I$,
\begin{eqnarray}
&&ds_4^2 = ds_2^2 + v_2(r,t) \, \left( d \theta^2 + \sin^2 \theta \, d \varphi^2 \right) \;, \nonumber\\
&& ds^2_2 = h_{ij} dx^i dx^j = dr^2 + {h}_{tt} (r,t) \, dt^2  \;, \nonumber\\
&&F_{rt}^I = e^I \;\;\;,\;\;\; F_{\theta \varphi}^I = p^I\, \sin \theta \;, \nonumber\\
&&{\hat A} = - 4 w^2 \;\;\;,\;\;\; X^I  \;,
\label{backgr}
\end{eqnarray}
where $X^I = X^I (r,t), \, w = w(r,t) $ and $e^I = e^I (r,t)$. Here we have written $ds^2_2$ in Fefferman-Graham form.
We will denote
\bea
\sqrt{-h} = \sqrt{ - \det h_{ij} } \;.
\eea
In the background \eqref{backgr}, the associated four-dimensional Ricci scalar $R_4$ is given by
\bea
R_4 &=&
R_2 -\frac{2}{v_2} + \frac{2}{v_2} \Box v_2 - \frac{\left(\partial v_2\right)^2 }{2 {v_2}^2 }\;,
\eea
where 
$R_2$ denotes the Ricci scalar associated to $ds_2^2$. Note that in the conventions used here \cite{Cardoso:2006xz}, the Ricci scalar of $S^2$ is negative.

Let us now discuss the evaluation of the four-dimensional Wilsonian Lagrangian  \eqref{eq:efflag}
in a background of the form \eqref{backgr} and its subsequent reduction on a two-sphere $S^2$. 
Following \cite{Cardoso:2006xz}, we introduce the rescaled fields
\begin{eqnarray}
   Y^I = \ft14 v_2 \, {\bar w} \, X^I \,,\quad
  \Upsilon = \ft{1}{16} v_2^2 \, {\bar w}^2 \, {\hat A} = - \ft14
  v_2^2 \, \vert w \vert^4  \;.
  \label{rescalXA}
  \end{eqnarray}
 Note that  $Y^I$ and $\Upsilon$ are $U(1)$ invariant, since $w$ and $X^I$ carry the same $U(1)$ weight.
 $\Upsilon$ is real and negative. In the rescaled variables, the relation \eqref{prep} becomes
 \bea
F(\lambda Y) = \lambda^2 F(Y) \;.
\label{FY}
\eea
The covariant derivative ${\cal D} X^I$ may be expressed as
\bea
\tfrac14 v_2 {\bar w} {\cal D}_{\mu}  X^I &=& \partial_{\mu} Y^I + i a_{\mu} Y^I - \tfrac12 \partial_{\mu} \Sigma  \, Y^I \;,
\label{covd}
\eea 
where we defined
\bea
\Sigma = \ln \left( v_2^2 |w|^2 \right) =  \ln \left( 2 \, v_2 \sqrt{-\Upsilon}  \right) \;,
\eea
and
where $a_{\mu}$ is given by the same expression as in \eqref{u1conn}, but with $X^I$ replaced everywhere by $Y^I$,
\bea
a_{\mu} = \tfrac12 \, e^{{\cal K} (Y)} \left( F_I (Y) \partial_{\mu} {\bar Y}^I - {\bar Y}^I \partial_{\mu} F_I (Y)+ c.c. \right) \;,
\label{u1connY}
\eea
where
	\be
	e^{- \mathcal{K}(Y)} = \, i \left( \bar{Y}^I F_I - \bar{F}_I Y^I   \right) \;.
	\label{calKY}
	\ee
Then, using \eqref{Pgc}, we infer the relation
\bea
v_2 \sqrt{-\Upsilon} = 8 G_4 \, e^{- \mathcal{K}(Y)} \;,
\label{vupsky}
\eea
which expresses $\Upsilon$ in terms of $Y^I$ and $v_2$, as well as the relation
\bea
\Sigma = \ln \left( 16 \, G_4 \, e^{- \mathcal{K}(Y)}  \right) \;.
\eea
Using \eqref{covd}, we obtain
\bea
&&\tfrac{1}{16} v_2^2 |w|^2 \,   i \left( {\cal D}^{\mu} F_I \, {\cal D}_{\mu} \bar X^I - c.c. \right) = 
 i \left( \partial^{\mu} F_I (Y) \, \partial_{\mu} \bar Y^I - \partial^{\mu} {\bar F}_I (\bar Y) \, \partial_{\mu}  Y^I \right) \nonumber\\
&&  -  e^{- {\cal K} (Y)} \left( a^{\mu}  a_{\mu} - \tfrac14 \partial^{\mu} \Sigma \partial_{ \mu} \Sigma \right) - \tfrac12 \partial^{\mu} \Sigma \,
\partial_{\mu} \left(    e^{- {\cal K}(Y)}\right) \;,
\label{XYconv}
  \eea
We then insert \eqref{XYconv} in \eqref{eq:efflag} to express the Lagrangian in terms of the scalar fields $Y^I, {\bar Y}^I$.

Subsequently, we evaluate the Lagrangian in the background \eqref{backgr}. Integrating over the two-sphere $S^2$ yields 
the reduced Lagrangian,
 \begin{eqnarray}
  \label{eq:F1-F2}
\ft12   \mathcal{F}_1 &=&{}  \tfrac18 N_{IJ} \Big[  ( \sqrt{-h}/v_2)^{-1}
  e^Ie^J - \frac{\sqrt{-h}}{v_2} \, {p^I} {p^J} \Big] -
  \tfrac14(F_{IJ}+ \bar F_{IJ}) {e^Ip^J}  \nonumber \\
  &&
  + \ft12 i e^I\Big[ F_I + F_{IJ} \bar Y^J   - \mathrm{h.c.}\Big] 
  - \ft12 \frac{\sqrt{-h}}{v_2} \,  p^I\Big[ F_I - F_{IJ} \bar Y^J  +  \mathrm{h.c.}\Big]\;,
  \nonumber\\[2mm] 
\ft12   \mathcal{F}_2 &=&  
    \frac{\sqrt{-h}}{2 G_4} \, 
  \left(1 - \ft12 v_2 \, \left(R_2 + \frac{2}{v_2} \Box v_2 - \frac{\left(\partial v_2 \right)^2 }{2 {v_2}^2 } \right) \right) \nonumber\\
  &&
  + i \frac{\sqrt{-h}}{v_2} \,  \Big[ F-Y^IF_I 
    +  \tfrac12 \bar
  F_{IJ}Y^IY^J - \mathrm{h.c.} \Big] \\
  &&+   \frac{ \sqrt{-h} \, v_2 \,  e^{ \mathcal{K}(Y)} }{2 G_4}  \Big( i \left( \partial^{\mu} F_I \, \partial_{\mu} \bar Y^I - \partial^{\mu} {\bar F}_I  \, \partial_{\mu}  Y^I \right) 
  \nonumber\\
&&  \qquad \qquad \qquad \qquad 
 -  e^{- {\cal K} (Y)} \left( a^{\mu}  a_{\mu} - \tfrac14 \partial^{\mu} \Sigma \partial_{ \mu} \Sigma \right) - \tfrac12 \partial^{\mu} \Sigma \,
\partial_{\mu} e^{- {\cal K}(Y)} \Big)
   \;, \nonumber
\end{eqnarray}
where now $F_I = \partial F(Y) / \partial Y^I, 
\, F_{IJ} = 
 \partial^2 F(Y) / \partial Y^I \partial Y^J$ and 
\begin{equation}
N_{IJ} = -i \left( F_{IJ} - \bar{F}_{IJ} \right) \;.
\label{nij}
\end{equation}

Next, to make the underlying electric-magnetic duality manifest, we perform a Legendre transformation  of $\cal F$ with respect to the $p^I$,
thereby replacing $p^I$ by the conjugate quantity $f_I = - \partial {\cal F} / \partial p^I $ \cite{Cardoso:2006xz}. 
Note that $f_I$ is a field strength in two dimensions, i.e.  $f_I = G_{I \, rt}$ with $G_I$ a two-form. 
The resulting Lagrangian $H(e^I, f_J) $,\footnote{We omit the dependence on the scalars $v_2, Y^I$, for notational simplicity.}
\begin{equation}
H(e^I, f_J) =  {\cal F} (e^I, p^J)  + p^I f_I \;,
\end{equation}
is the two-dimensional bulk Lagrangian that we will work with.
In order to make the electric-magnetic duality of $H(e^I, f_J) $ manifest, we define
\bea
R_{IJ} = F_{IJ} + {\bar F}_{IJ} \;\;\;,\;\;\; N^{IJ} N_{JK} = \delta^I_K \;.
\eea
Then, upon dropping  a total derivative term, 
\begin{align}  
\label{lagHef}
H(e^I, f_J) 
=& \; \frac14 \, ( \sqrt{-h}/v_2)^{-1} \, (e^I, f_I)
\begin{bmatrix}
N_{IJ} + R_{IK} N^{KL} R_{LJ}  & \;\;\; - 2 R_{IK} N^{KJ} \\
- 2 N^{I K} R_{KJ} & 4 N^{IJ}
\end{bmatrix} \, 
\begin{pmatrix}
e^J \\ 
f_J
\end{pmatrix}
\nonumber\\
&\; +  2i \  (e^I, f_I) 
\begin{pmatrix}
F_I - {\bar F}_I \\
- (Y^I - {\bar Y}^I) 
\end{pmatrix} 
+\sqrt{-h}\,  \frac{  v_2  }{2 G_4 } \,
\, \left[ - R_2 +\frac{2}{v_2} +  \frac{\left(\partial v_2\right)^2}{2 {v_2}^2 }  \right]  \nonumber\\
& \;   -\sqrt{-h}\;  \frac{ 2\, e^{-\mathcal{K} (Y)} }{v_2}  \\
&\; -   \frac{\sqrt{-h}\;  v_2 \,  e^{\mathcal{K} (Y)}  }{ G_4  } \left[  N_{IJ}  \partial  Y^I \, \partial  \bar Y^J    +  e^{\mathcal{K}(Y)}\, N_{IJ} Y^I \partial \bar{Y}^J\; N_{KL} \bar{Y}^L \partial Y^K  \right]   \,. \nonumber
\end{align} 
When taking the complex scalar fields $Y^I$  to be constant, this two-dimensional bulk Lagrangian describes a dilaton-gravity action in two
dimensions that reduces to the one in \cite{Sarosi:2017ykf} when considering the reduction of the four-dimensional Einstein+Maxwell theory on a two-sphere.

\section{Equations of motion}

The two-dimensional bulk action \eqref{lagHef}  gives rise to the following equations of motion.
The equations of motion for the field strengths $e^I$ and $f_J$ are solved by \cite{Aniceto:2020saj}
\be
\begin{pmatrix}
 e^I\\
f_I
\end{pmatrix}  = \frac{\sqrt{-h}}{2 v_2}   \left\{   \begin{bmatrix}
2 N^{IK}R_{KJ}   & -4 N^{IJ}  \\ 
N_{IJ} + R_{IK}N^{KL} R_{LJ} & -2 R_{IK} N^{KJ}
\end{bmatrix}   \begin{pmatrix}
p^J \\ q_J
\end{pmatrix}      +4 \begin{pmatrix}
Y^I + \bar{Y}^I \\ 
F_I + \bar{F}_I
\end{pmatrix}     \right\} \,, \label{eq_of_motion_eI_fI}
\ee
with integration constants $p^I$ and $q_I$.

In what follows, we restrict to linearized order in fluctuations and in units of $G_4=1$.
The equation of motion for $Y^I$ reads,
\bea
\Delta_I - 2 \partial_I e^{-{\cal K}(Y)} + v_2^2  \, e^{{\cal K}(Y)} \left( N_{IJ} \Box {\bar Y}^J + e^{{\cal K}(Y)} N_{IJ} {\bar Y}^J \, N_{KL} Y^K \Box {\bar Y}^L 
\right) = 0 \;,
\label{eomY}
\eea
where
\bea
\partial_I e^{-{\cal K}(Y)} &=& -  N_{IJ} {\bar Y}^J \;, \nonumber\\
\Delta_M &=& 
  i \bar{F}_{IK} N^{KL} F_{LPM} \left(2 N^{PJ} q_J -    N^{PQ}   \bar{F}_{QJ}\,    p^J    \right)p^I  - i N^{IK}F_{KLM} N^{LJ} \,q_I q_J  \nonumber\\
 && +2 q_M -2 F_{IM} \, p^I -4 N_{IM} \bar{Y}^I  \;.
 \eea
 In the case of solutions with constant $Y^I$, the resulting algebraic equations are the attractor equations given in \cite{Cardoso:2006xz}.

 Using
 \bea
 \delta R_2 = \left( h^{ij} \Box - \nabla^{i} \nabla^{j} \right) \delta h_{ij} - R^{ij} \delta h_{ij} \;,
 \eea
we obtain for the equation of motion for the  two-dimensional metric $ds_2^2 = h_{ij} dx^{i} dx^{j}$,
\bea
&& h^{ij} \left[
- \frac18 \, \frac{v_2}{
(\sqrt{-h})^2} \, (e^I, f_I)
\begin{bmatrix}
N_{IJ} + R_{IK} N^{KL} R_{LJ}  & \;\;\; - 2 R_{IK} N^{KJ} \\
- 2 N^{I K} R_{KJ} & 4 N^{IJ}
\end{bmatrix} \, 
\begin{pmatrix}
e^J \\ 
f_J
\end{pmatrix} + \frac12 - \frac{e^{- {\cal K}(Y)}}{v_2}  - \frac12 \Box v_2
\right] \nonumber\\
&& + \frac12 \nabla^{i} \nabla^{j} v_2 = 0 \;.
\label{hfluc}
\eea
Taking the trace of this equation gives
\bea
\label{thfluc}
- \frac14 \, \frac{v_2}{
(\sqrt{-h})^2} \, (e^I, f_I)
\begin{bmatrix}
N_{IJ} + R_{IK} N^{KL} R_{LJ}  & \;\;\; - 2 R_{IK} N^{KJ} \\
- 2 N^{I K} R_{KJ} & 4 N^{IJ}
\end{bmatrix} \, 
\begin{pmatrix}
e^J \\ 
f_J
\end{pmatrix} + 1 - 2 \frac{e^{- {\cal K}(Y)}}{v_2}  - \frac12 \Box v_2
= 0 \;. \nonumber\\
\eea
The equation of motion for $v_2$ reads
\begin{align}  
\label{vfluc}
& \; \frac14 \, \frac{1}{( \sqrt{-h})^2 } \, (e^I, f_I)
\begin{bmatrix}
N_{IJ} + R_{IK} N^{KL} R_{LJ}  & \;\;\; - 2 R_{IK} N^{KJ} \\
- 2 N^{I K} R_{KJ} & 4 N^{IJ}
\end{bmatrix} \, 
\begin{pmatrix}
e^J \\ 
f_J
\end{pmatrix}
\nonumber\\
&\; - \frac12 R_2 -   \frac12  \frac{\Box v_2}{ v_2}    +   \frac{ 2\, e^{-\mathcal{K}(Y)} }{v_2^2} = 0 \;.
\end{align} 
Dividing \eqref{thfluc} by $v_2$ and adding \eqref{vfluc} to it gives
\bea
R_2 = 2 \left( \frac{1}{v_2} -  \frac{\Box v_2}{ v_2} \right) \;.
\label{flucmet}
\eea

\subsection{Constant curvature solution}

The above equations of motion admit the following constant curvature solution supported by constant scalar fields $v_2$ and $Y^I$ \cite{Aniceto:2020saj},
\bea
ds^2_2 &=& dr^2 + {h}_{tt} (r,t) \, dt^2  \;\;\;,\;\;\; - {h}_{tt} =\left(  {\alpha} (t) \, e^{r/\sqrt{v_1}} + {\beta} (t) \, e^{-r/\sqrt{v_1}} \right)^2 \;,
 \nonumber\\
  v_1 &=& v_2 = e^{- {\cal K} (Y)} \, G_4 
  =  i (\bar Y^IF_I -Y^I\bar F_I ) \, G_4
   \;,  \nonumber\\
Y^I - {\bar Y}^I &=& i p^I \;\;\;,\;\,\; F_I - {\bar F}_I = i q_I  \;, \nonumber\\
\begin{pmatrix}
f_I \\
e^I
\end{pmatrix} &=& \frac{ \sqrt{-h} }{v_2}  
\begin{pmatrix}
F_I +{\bar F}_I  \\
Y^I + {\bar Y}^I 
\end{pmatrix} \;.
 \label{2dBPS}
\eea
The two-dimensional space-time metric has constant curvature scalar, $R_2 = 2/v_1 >0$, where
$v_1$ denotes a constant scale factor of length dimension 2. 
The scalar fields $Y^I$ satisfy the horizon BPS attractor equations for single centre BPS black holes in four dimensions \cite{LopesCardoso:2000qm}.
Expressing 
the field strengths $e^I = F_{rt}^I $ and $f_I = G_{I \, rt} $ in terms of gauge potential one-forms, 
\bea
e^I = F_{rt}^I = \partial_r A_t^I - \partial_t A_r^I \;\;\;,\;\;\; f_I = G_{I \, rt} = \partial_r \tilde{A}_{I \, t} - \partial_t \tilde{A}_{I \, r} \;,
\eea
and working in the gauge 
\bea
A^I_r =  \tilde{A}_{I \, r}  = 0 \;,
\label{rgA}
\eea
the gauge potentials associated with the field strengths \eqref{eq_of_motion_eI_fI} take the following form in this background \cite{Aniceto:2020saj},
 \bea 
  \begin{pmatrix}
\tilde{A}_{t I  }\\
- A_t^I
\end{pmatrix} = \sqrt{v_1} \,  \frac{ 
{\alpha} (t) \, e^{r/\sqrt{v_1}}  }{ {\sqrt{-h}} }
\left( 1 - \frac{ \beta (t)}{\alpha (t) } \,  e^{-2r/\sqrt{v_1}} \right)
\begin{pmatrix}
f_I \\
- e^I
\end{pmatrix} + \begin{pmatrix}
\tilde{\mu}_I (t) \\
- \mu^I (t)
\end{pmatrix} \;.
\label{AtAr}
\eea

Setting $ \sqrt{v_1} \, \alpha(t) = 1, \, \sqrt{v_1} \, \beta (t) = - \mu^2/4$ in \eqref{2dBPS}, with constant $\mu^2$, yields a two-dimensional
solution describing a black hole in $AdS_2$ \cite{Spradlin:1999bn}. We will return to this solution in section \ref{sec:RN}.

\subsection{First-order perturbations around constant curvature solution \label{sec:flucana}}

In the following, we will take the solution \eqref{2dBPS} as a background and 
consider perturbations around this background, 
\bea
\sqrt{-h} &=& \sqrt{-h_0} + \sqrt{-h_1} \;, \nonumber\\
v_2 &=& v_{2,0} + \chi \;, \nonumber\\
Y^I &=& Y^I_0 + {\cal Y}^I \;,
\label{fopert}
\eea
where $ \sqrt{-h_0}, v_{2,0}, Y^I_0$ will denote the background expressions \eqref{2dBPS}.
The field strengths $e^I$ and $f_J$ are then perturbed according to \eqref{eq_of_motion_eI_fI}.
We will work to first order in the perturbations $\sqrt{-h_1}, \chi, {\cal Y}^I $.  At this order,
we may drop terms in the equations of motion
 that involve more than one field being
differentiated, since these terms will be of higher order in the perturbations.
To keep track of the order of the perturbation, 
we will attach a parameter $\epsilon$
to each of above perturbations.

To first order in $\epsilon$, the equation of motion \eqref{eomY} becomes
\bea
N^{(0)}_{IJ} \left( \Box - \frac{2}{v_1} \right) {\bar{\cal Y}}^J + e^{ {\cal K}(Y_0)} N_{IJ} {\bar Y}^J_ 0 \, N_{KL} Y^K_0 \Box {\bar {\cal Y}}^L = 0 \;,
\label{Yfluc}
\eea
where $N^{(0)}_{IJ}$ denotes the combination \eqref{nij}  evaluated at $Y^I_0$.
Contracting this expression with $Y^I_0$ gives
\bea
Y^I_0 N^{(0)}_{IJ} \, {\bar {\cal Y} }^J = 0 \;,
\label{YdY}
\eea
which implies that the first-order perturbation of $e^{ -{\cal K}(Y)}$ vanishes,
\bea
\delta e^{- {\cal K}(Y)} = 0 \;.
\eea
Thus, \eqref{Yfluc} reduces to
\bea
 \left( \Box - \frac{2}{v_1} \right) {\bar{\cal Y}}^I = 0 \;.
 \label{pdeY}
 \eea
 We note that 
 the condition \eqref{flucmet} becomes
 \bea
\frac{1}{\sqrt{-h_0}} \left( \partial_r^2 - \frac{1}{v_1} \right) \sqrt{-h_1} = - \frac{\chi}{v_1^2} - \frac{\Box \chi}{v_1} \;.
\label{eqfluch}
\eea

Next, we use \eqref{eq_of_motion_eI_fI} to obtain
\bea
 && (e^I, f_I)
\begin{bmatrix}
N_{IJ} + R_{IK} N^{KL} R_{LJ}  & \;\;\; - 2 R_{IK} N^{KJ} \\
- 2 N^{I K} R_{KJ} & 4 N^{IJ}
\end{bmatrix} \, 
\begin{pmatrix}
e^J \\ 
f_J
\end{pmatrix} \nonumber\\
&&
= \left( \frac{\sqrt{-h}}{v_2} \right)^2 \left[
 \left(p^I , q_I \right)      \begin{bmatrix}
	N_{IJ} + R_{IK}N^{KL}R_{LJ} & -2 R_{IK} N^{KJ} \\ 
	-2 N^{IK} R_{KJ} & 4 N^{IJ} 
\end{bmatrix}         \begin{pmatrix}
	p^J \\ q_J
\end{pmatrix}  \right.    \nonumber  \\
&& \left. \qquad \qquad   + 8 \left( Z(Y) + {\bar Z} ({\bar Y}) \right)  
  -16 e^{-\mathcal{K}(Y)}
\right] \;,
\label{efpq}
\eea
where
\bea
Z(Y) = p^I F_I  (Y) - q_I Y^I  \;.
\label{Zqp}
\eea
We note that under a first-order perturbation around the background \eqref{2dBPS},
\begin{align}
\label{flucmax}
&\delta  \Bigg\{  \tfrac14  \left(p^I , q_I \right)      \begin{bmatrix}
N_{IJ} + R_{IK}N^{KL}R_{LJ} & -2 R_{IK} N^{KJ} \\ 
-2 N^{IK} R_{KJ} & 4 N^{IJ} 
\end{bmatrix}         \begin{pmatrix}
p^J \\ q_J
\end{pmatrix}     + 2 \left(p^I , q_I\right) \begin{pmatrix}
F_I + \bar{F}_I \\ - \left(Y^I +\bar{Y}^I\right)
\end{pmatrix}  \Bigg\} = \nonumber \\
& \quad -2 N^{(0)}_{IJ} \left( Y^I_0  \bar{\mathcal{Y} }^J + \bar{Y}^I_0 \mathcal{Y}^J  \right) = 2 \,  \delta e^{-\mathcal{K}(Y)} = 0\,.
\end{align}
Using this equation, we obtain
\bea
\label{flpqz}
\delta  \left\{
 \left(p^I , q_I \right)      \begin{bmatrix}
	N_{IJ} + R_{IK}N^{KL}R_{LJ} & -2 R_{IK} N^{KJ} \\ 
	-2 N^{IK} R_{KJ} & 4 N^{IJ} 
\end{bmatrix}         \begin{pmatrix}
	p^J \\ q_J
\end{pmatrix} 
  + 8 \left( Z(Y) + {\bar Z} ({\bar Y} ) \right)  
  -16 e^{-\mathcal{K}(Y)}
\right\} = 0 \;.\nonumber\\
\eea
Next, using this result in  \eqref{vfluc}, we obtain
\bea
\frac{1}{\sqrt{-h_0}} \left( \partial_r^2 - \frac{1}{v_1} \right) \sqrt{-h_1} = - 2 \frac{\chi}{v_1^2} - \frac12 \frac{\Box \chi}{v_1} \;.
\label{v2cons}
\eea
Comparing this equation with  \eqref{eqfluch}  we infer
\bea
\Box \chi = 2 \frac{\chi}{v_1} \;.
\label{eqrch}
\eea
Inserting this back into  \eqref{v2cons} gives
\bea
\frac{1}{\sqrt{-h_0}} \left( \partial_r^2 - \frac{1}{v_1} \right) \sqrt{-h_1} = - 3 \frac{\chi}{v_1^2} \;.
\label{diffmefl}
\eea

Now we consider the equation of motion \eqref{hfluc}, which we rewrite, using \eqref{efpq},
as
\bea
&& h^{ij} \left\{
-  \frac{1}{v_2}
 \left[ \frac18
 \left(p^I , q_I \right)      \begin{bmatrix}
	N_{IJ} + R_{IK}N^{KL}R_{LJ} & -2 R_{IK} N^{KJ} \\ 
	-2 N^{IK} R_{KJ} & 4 N^{IJ} 
\end{bmatrix}         \begin{pmatrix}
	p^J \\ q_J
\end{pmatrix}  \right.  \right.  \nonumber  \\
&& \left. \left. \qquad \qquad   +  \left( Z(Y) + {\bar Z} ({\bar Y}) \right)  
  - 2 e^{-\mathcal{K}(Y)}
\right] 
+ \frac12 - \frac{e^{- {\cal K}}}{v_2}  - \frac12 \Box v_2
\right\} \nonumber\\
&& + \frac12 \nabla^{i} \nabla^{j} v_2 = 0 \;.
\label{metflhmn}
\eea
Using \eqref{flpqz}, the $rr$-component of this equation gives rise to
\bea
\partial_r^2 \chi - \Box \chi + \frac{\chi}{v_1} = 0 \;.
\label{rreomd}
\eea
Combining with \eqref{eqrch} gives
\bea
\partial_r^2 \chi =  \frac{\chi}{v_1} \;.
\label{harchi}
\eea
Equation \eqref{harchi} is solved by
\bea
\chi = \nu(t) \, e^{r/\sqrt{v_1}} + \vartheta (t) \, e^{- r/\sqrt{v_1}} \;.
\label{solchi}
\eea
Since $\chi$ is of order $\epsilon$, so are $\nu$ and $\vartheta$.

The $tr$-component of \eqref{metflhmn} gives rise to 
\bea
\partial_t \partial_r \chi - \left( \partial_r \log \sqrt{- h_0} \right) \, \partial_t \chi = 0 \;.
\eea
Inserting \eqref{solchi} into this equation gives rise to three equations that are obtained by equating the exponential terms.
Two of these equations are identically satisfied.  The third equation gives the relation 
\bea
\nu' \beta = \vartheta' \alpha \;,
\label{bnv}
\eea
where $' = \partial_t$.
There are two cases to consider: either $\nu' =0$ or  $\nu'  \neq 0$. If  $\nu' =0$ then 
$\vartheta' =0$ (if we assume that $\alpha \neq 0$). 
On the other hand, if $\nu'  \neq 0$, then
\bea
\beta = \alpha \, \frac{\vartheta'}{\nu'} \;.
\label{betthe}
\eea
Thus, $\beta$ is now determined in terms of $\alpha, \nu$ and $\vartheta$.
Since both $\nu$ and $\vartheta$ are of order $\epsilon$, $\beta$ is of order $\epsilon^0$, as required for the background \eqref{2dBPS}.

The $tt$-component of \eqref{metflhmn} gives rise to 
\bea
\nabla^t \nabla_t \chi + \frac{\chi}{v_1} - \Box \chi = 0 \;.
\label{httflu}
\eea
Expressing $\Box = \nabla^t \nabla_t + \nabla^r \nabla_r = \nabla^t \nabla_t + \partial_r^2$, we see that \eqref{httflu} equals
\eqref{harchi}. On the other hand, inserting \eqref{eqrch} into \eqref{httflu} gives
\bea
\nabla^t \nabla_t \chi - \frac{\chi}{v_1}  = 0 \;,
\eea
which equals \eqref{rreomd}.
Writing out this equation gives
\bea
\partial_t^2 \chi - \left( \partial_t \log \sqrt{- h_0} \right) \partial_t \chi + h_{tt,0} \left( \left( 
\partial_r \log \sqrt{- h_0} \right) \partial_r \chi - \frac{\chi}{v_1} \right) = 0 \;.
\label{2th}
\eea
Inserting
\bea
\sqrt{- h_0} = \alpha(t) \, e^{r /\sqrt{v_1}} + \beta(t) \, e^{- r /\sqrt{v_1}} 
\label{met0}
\eea
as well as \eqref{solchi} into \eqref{2th}, we obtain three equations that are obtained by equating the exponential terms, namely
\bea
\nu'' \alpha - \nu' \alpha' + \frac{2}{v_1} (\alpha \vartheta + \beta \nu) \alpha^2 &=& 0 \;, \nonumber\\
\nu'' \beta + \vartheta'' \alpha - \vartheta' \alpha' - \nu' \beta' +  \frac{4}{v_1} (\alpha \vartheta + \beta \nu) \alpha \beta &=& 0  \;, \nonumber\\
\vartheta'' \beta - \vartheta' \beta' + \frac{2}{v_1} (\alpha \vartheta + \beta \nu) \beta^2 &=& 0 \;.
\label{nvab}
\eea
There are several cases to consider. We will focus on the following two cases. In the first case we take $\nu' = \vartheta' = 0$, in which case we
infer (assuming $\alpha \neq 0$)
\bea
\alpha \vartheta + \beta \nu = 0 \;.
\label{abnv0}
\eea
In the second case we take $ \nu' \neq 0, \vartheta' \neq 0$, in which case we have the relation \eqref{betthe}. 
In the following, we will focus on this second case.
Let us consider the last equation in \eqref{nvab}. Using \eqref{betthe}, we obtain (assuming $\alpha \neq 0$)
\bea
\vartheta' + \frac{\nu'}{\nu} \vartheta = - \frac{v_1}{2} \frac{ \nu'}{\alpha \nu} \left( \frac{\nu'}{\alpha} \right)' \;.
\label{odeth}
\eea
This ODE for $\vartheta$ is solved by
\bea
\vartheta = \frac{1}{\nu} \left( c_0 - \frac{v_1}{4}  \left( \frac{\nu'}{\alpha} \right)^2 \right) \;.
\label{solthe}
\eea
Since both $\nu$ and $\vartheta$ are of order $\epsilon$, the integration constant $c_0$ is of order $\epsilon^2$. Using \eqref{odeth} and 
\eqref{solthe} in \eqref{bnv}, we infer the relation
\bea
\beta \nu = \alpha \frac{\nu}{\nu'} \, \vartheta' = - \alpha \vartheta - \frac{v_1}{2} \left( \frac{\nu'}{\alpha}\right)' 
= - \frac{\alpha}{\nu} \left( c_0 - \frac{v_1}{4}  \left( \frac{\nu'}{\alpha} \right)^2 \right) - \frac{v_1}{2} \left( \frac{\nu'}{\alpha}\right)' \;.
\label{schwarzcal}
\eea

Next, we find that the first two equations in \eqref{nvab} are identically satisfied, as can be verified by using \eqref{betthe} and 
\eqref{odeth}. Thus, there are no constraints on $\nu(t)$ other than $\nu' \neq 0$, which we assumed in the above. $\vartheta$ is specified by 
\eqref{solthe}, while $\beta $ has to satisfy \eqref{betthe}.

Finally, we return to  \eqref{diffmefl}  and solve for the 
metric fluctuation $\sqrt{-h_1}$,
\be
\sqrt{-h_1} = -\frac{\sqrt{-h_0}}{v_1} \, \chi - 2\, \partial_t \left(  \frac{\partial_t \nu }{\alpha}\right)  \,,
\label{h1p}
\ee
up to a homogeneous solution that can be absorbed into the definitions of $\alpha$ and $\beta$ \cite{Castro:2018ffi}.
Thus,
\bea
\sqrt{-h} = \sqrt{-h_0} + \sqrt{-h_1} = \sqrt{-h_0} \left( 1 - \frac{\chi}{v_1} \right) - 2\, \partial_t \left(  \frac{\partial_t \nu }{\alpha}\right)  \,.
\label{hhcm}
\eea

To ensure that the perturbations $\chi$ and $\sqrt{-h_1}$ are small, we demand \cite{Castro:2018ffi}
\bea
\big\vert \frac{\chi}{v_1} \big\vert \ll 1
\label{magnfluc}
\eea
which, for large $r$ becomes
\bea
\big\vert \frac{\nu}{v_1} \, e^{r/\sqrt{v_1}}  \big\vert \ll 1 \;.
\label{condnu}
\eea
This can be ensured \cite{Castro:2018ffi}
by introducing a cutoff $r_c$ for the radial coordinate $r$ at large $r$, 
$r_c \gg 1$, and by demanding that the amplitude
$\nu$ is accordingly small, so that \eqref{condnu} is satisfied for a large cutoff $r_c$.
Under this assumption, the asymptotic behaviour of $\sqrt{-h}$ is given by the asymptotic behaviour of $\sqrt{-h_0} $,
which is what we will use.

The wave equations \eqref{pdeY} and \eqref{eqrch} for the scalar perturbations are massive Klein-Gordon equations with mass-square parameter
$m^2 = 2/v_1$. These perturbations correspond to boundary operators with scaling dimension $\Delta =2$, 
\bea
\Delta = \frac12 \left( d + \sqrt{d + 4 m^2 v_1} \right) = 2 \;,
\eea
where $d=1, m^2 = 2/v_1$.
Since $\Delta > d=1$, these operators are irrelevant operators \cite{Cvetic:2016eiv,Castro:2018ffi,Castro:2019vog}.

Thus, summarizing, at order $\epsilon$ we have the perturbations
\bea
\sqrt{-h_1} &=& -\frac{\sqrt{-h_0}}{v_1} \, \chi - 2\, \partial_t \left(  \frac{\partial_t \nu }{\alpha}\right)  \;, \nonumber\\
\chi &=& \nu(t) \, e^{r/\sqrt{v_1}} + \vartheta (t) \, e^{- r/\sqrt{v_1}} \;, \nonumber\\
 \left( \Box - \frac{2}{v_1} \right) {\bar{\cal Y}}^I &=& 0 \;\;\;,\;\;\; Y^I_0 N^{(0)}_{IJ} \, {\bar {\cal Y} }^J = 0 \;.
 \label{pertsol}
\eea
These induce a perturbation of 
the field strengths $e^I$ and $f_J$ according to \eqref{eq_of_motion_eI_fI}.
The perturbations $\nu$ and $\vartheta$ satisfy
additional conditions, as discussed above. For instance, when $\nu' = \vartheta' = 0$, $\nu$ and $\vartheta$ have to 
satisfy the condition (assuming $\alpha \neq 0$)
\bea
\alpha \vartheta + \beta \nu = 0 \;,
\label{avbn2}
\eea
while when $\nu' \neq 0, \vartheta' \neq 0$, 
they satisfy the relations (assuming $\alpha \neq 0$)
\bea
\vartheta &=& \frac{1}{\nu} \left( c_0 - \frac{v_1}{4}  \left( \frac{\nu'}{\alpha} \right)^2 \right) \;, \nonumber\\
\beta &=& \alpha \, \frac{\vartheta'}{\nu'} = - \frac{\alpha}{\nu^2} \left( c_0 - \frac{v_1}{4}  \left( \frac{\nu'}{\alpha} \right)^2 \right) - \frac{v_1}{2\, \nu} \left( \frac{\nu'}{\alpha}\right)' \;,
\label{pertsol2}
\eea
where the integration constant $c_0$ is of order $\epsilon^2$. 
The last relation in \eqref{pertsol2} can also be expressed as
\bea
\alpha \vartheta + \beta \nu = - \frac{v_1}{2} \left( \frac{\nu'}{\alpha}\right)' \;.
\label{avbnmod}
\eea
These results are the counterpart of those obtained previously in a different context, namely
five-dimensional pure Einstein gravity with cosmological constant
reduced to two dimensions \cite {Castro:2018ffi}, 
and three-dimensional topologically massive gravity reduced to two dimensions \cite{Castro:2019vog}.

In the following, we will set ${\cal Y}^I =0$.  
We will thus only keep the perturbation $\chi$ given in \eqref{solchi}, which then induces the 
perturbations of the remaining fields as described above. 
In section \ref{sec:calogero} we will discuss the boundary action that describes the dynamics of $\nu (t)$. 
This boundary action 
is the DFF action \cite{deAlfaro:1976vlx}. In \cite{Gibbons:1998fa}, the $n$ particle extension of the DFF model has been proposed as a quantum mechanics
model for the microstates of the extremal Reissner-Nordstrom in four dimensions.

Next, we will recast double-extreme
and near-extremal four-dimensional black holes in the language of the two-dimensional 
perturbation analysis given above.

\subsubsection{Double-extreme four-dimensional BPS black hole solutions \label{sec:de}}

In the following we will consider the dimensional reduction of double-extreme four-dimensio\-nal single centre BPS black hole solutions. These four-dimensional solutions
have a near-horizon geometry given by $AdS_2 \times S^2$.  By working one step away from the horizon, 
we will recast these interpolating solutions in terms of the perturbation analysis performed
in the previous subsection.

In \cite{LopesCardoso:2000qm}, 
the four-dimensional BPS black hole solutions were given in terms of rescaled fields $Y^I$ and $\Upsilon$. However,
the rescaled fields used in \cite{LopesCardoso:2000qm} differ from the rescaled fields \eqref{rescalXA} used here. We will therefore denote the fields
$(Y^I, \Upsilon)$ used in  \cite{LopesCardoso:2000qm} with a hat, i.e. $({\hat Y}^I, \hat{\Upsilon})$, so as to distinguish them from the fields defined
in \eqref{rescalXA}. In Appendix \ref{relYU} we will determine how these two sets of fields are related, and this relation will be used
below.

We write the 
four-dimensional space-time metric of the four-dimensional single centre BPS black hole solution
as in \cite{LopesCardoso:2000qm},
\bea
ds_4^2 = - e^{2g(R)} \, dt^2 + e^{- 2 g(R)} \, \left( dR^2 + R^2  \, \left( d \theta^2 + \sin^2 \theta \, d \varphi^2 \right) \right)\;.
\label{lin4dR}
\eea
The metric factor $e^{- 2 g} $ is determined in terms of the scalar fields ${\hat Y}^I$ by
\bea
e^{- 2 g} =G_4 \,  i \left( \bar{\hat{Y}}^I \, F_I (\hat{Y}) - {\bar F}_I (  \bar{\hat{Y}} ) \, \hat{Y}^I \right) \;,
\label{egY}
\eea
where the ${\hat Y}^I$ are determined in terms of harmonic functions $(H^I, H_I)$ by 
\bea
\begin{pmatrix} 
{\hat Y}^I - \bar{\hat{Y}}^I \\
F_I (\hat{Y}) - {\bar F}_I (  \bar{\hat{Y}} ) 
\end{pmatrix}
= i 
\begin{pmatrix}
H^I \\
H_I
\end{pmatrix} =  i 
\begin{pmatrix}
h ^I + \frac{p^I}{R} \\
h_I + \frac{q_I}{R} 
\end{pmatrix} \;,
\label{hatYHH}
\eea
where $(h^I, h_I) \in \mathbb{R}^2$ denote integration constants, subject to 
$h^I q_I - h_I p^I = 0$.
We recall the BPS flow equation (c.f. \eqref{floww})
\bea
R^2 \, \partial_R g = G_4 \, e^{2g} \left(p^I \,  {F}_I ({\hat Y})   - 
q_I \,   {\hat Y}^I \right) \;,
\label{flowpg}
  \eea
  where $\partial_R g > 0$ \cite{Ferrara:1997tw}.
The metric factor $v_2$  in \eqref{backgr} is given by
\bea
v_2 = e^{- 2g} \, R^2 
= G_4 \, R^2 \, \left(  H^I  {F}_I ({\hat Y})  - H_I   {\hat Y}^I \right) \;,  
\label{relYY}
\eea
where we used \eqref{hatYHH}. The horizon is located at $R=0$.

Using the relations given in Appendix \ref{relYU}, 
\bea
Y^I = {\hat Y}^I \,  \frac{ R^2 \, \sqrt{- \hat{\Upsilon}} }{  8 } \;\;\;,\;\;\; \Upsilon = - {\hat \Upsilon}^2  \frac{R^4 }{64} \;,
\label{Yupsrel2}
\eea
as well as \cite{LopesCardoso:2000qm}
\bea
\hat{\Upsilon} = - 64 \, \left( \partial_R g \right)^2 \;,
\label{upsg}
\eea
we obtain 
\bea 
Y^I =  {\hat Y}^I \,   R^2 \ | \partial_R g |   \;\;\;,\;\;\;
\Upsilon = -  64  \, R^4  \left( \partial_R g \right)^2 \;.
\label{yyuu}
\eea

Next, we specialise to double-extreme single centre BPS black holes, i.e. BPS black holes 
supported by constant scalar fields $z^A = Y^A/Y^0$ (with $A = 1, \dots, n$). To this end, we set
$h^I = c \, p^I, h_I = c \, q_I$, 
with $c \in \mathbb{R}^+$. Using \eqref{hatYHH}, we obtain
\bea
\begin{pmatrix} 
{\hat Y}^I - \bar{\hat{Y}}^I \\
F_I (\hat{Y}) - {\bar F}_I (  \bar{\hat{Y}} ) 
\end{pmatrix}
= i 
\begin{pmatrix}
p^I \\
q_I
\end{pmatrix} \, f(R) \;\;\;,\;\;\; f(R) = c + \frac{1}{R} \;.
\eea
Solving these equations gives rise to
\bea
{\hat Y}^I  = A^I \, f(R) \;,
\eea
where the constants $A^I$ denote the solutions to the horizon attractor equations
\bea
\begin{pmatrix} 
{\hat Y}^I - \bar{\hat{Y}}^I \\
F_I (\hat{Y}) - {\bar F}_I (  \bar{\hat{Y}} ) 
\end{pmatrix}
= i 
\begin{pmatrix}
p^I \\
q_I
\end{pmatrix} \;.
\eea
Then, from \eqref{relYY} and \eqref{flowpg} we infer
\bea
v_2 &=& G_4 \, \left( R \, f(R) \right)^2 \, \left(  p^I  {F}_I (A)  - q_I   A^I \right) \;, \nonumber\\
R^2 \, \partial_R g &=& \frac{1}{f(R)} \;,
\eea
and hence, using \eqref{yyuu} we obtain
\bea
Y^I = A^I \;\;\;,\;\;\; \Upsilon = - \frac{64 }{f^2(R)} \;.
\eea
Thus, for double-extreme single centre BPS black holes, the scalar fields $Y^I$ are constant. 
The four-dimensional line element \eqref{lin4dR} then takes the form
\bea
ds_4^2 = - \frac{R^2}{ v_{2,0} \, ( c R + 1)^2} \, dt^2 + v_{2,0} \, \frac{( c R + 1)^2}{R^2} \, dR^2 + v_{2,0} \, (c R + 1)^2 \, 
\left( d \theta^2 + \sin^2 \theta \, d \varphi^2 \right) \;,\nonumber\\
\label{dem}
\eea
where
\bea
v_{2,0} = G_4 \,  \left(  p^I  {F}_I (A)  - q_I   A^I \right) \;.
\eea
When $c = 0$, this describes an $AdS_2 \times S^2$ solution supported by constant scalars $Y^I$.

Next we convert the radial coordinate $R$ to the radial coordinate $r$ in \eqref{backgr} using $dr = e^{-g(R)} \, dR$, 
\bea
\frac{r}{\sqrt{v_1}} = \log R + c R \;,
\eea
where $v_1 = v_{2,0}$. Working away from the horizon at order $c R \ll 1$, we infer
\bea
R = e^{r/\sqrt{v_1}} - c  \,  e^{2r/\sqrt{v_1}} \;.
\eea
At this order, the metric \eqref{dem} reads
\bea
ds_4^2 = - \frac{e^{2r/\sqrt{v_1}}}{ v_{2,0}} \, \left( 1 - 4 c  \, e^{r/\sqrt{v_1}} \right) \, dt^2 +  dr^2 + v_{2,0} \, \left(1 + 2 c \, e^{r/\sqrt{v_1} }\right) \, 
\left( d \theta^2 + \sin^2 \theta \, d \varphi^2 \right) \;. \nonumber\\
\eea
It can be viewed as a perturbation of an $AdS_2 \times S^2$ background using \eqref{fopert},
\bea
{\cal Y}^I &=& 0 \;,
\nonumber\\
{\cal \chi} &=& 2 c \, v_{2,0} \, e^{r/\sqrt{v_1}} \;, \nonumber\\
\sqrt{- h_1} &=& - 2 c \,  \frac{e^{2r/\sqrt{v_1}}}{ \sqrt{v_{2,0}}} = - \frac{ \sqrt{- h_0} }{v_1} \, \chi \;.
\label{nads}
\eea
This is of the form \eqref{pertsol} with constant $\nu = 2 c \, v_1 $ and
vanishing $\vartheta$, and satisfies \eqref{abnv0}.

The above perturbation can be viewed as the infinitesimal form of a symmetry transformation, called Harrison transformation, which transforms the double-extreme 
near-horizon geometry to a double-extreme interpolating solution, keeping the scalar fields at the attractor values \cite{Goldstein:2014gta}.

\subsubsection{Near-extremal Reissner-Nordstrom solutions \label{sec:RN}}

We describe how to view a 
four-dimensional near-extremal Reissner-Nordstrom black hole as a specific deformation of a 
 two-dimensional black hole in an $AdS_2$ space-time by means of an infinitesimal Harrison transformation.
We set  $G_4 = 1$ for simplicity in the following.

We will consider near-extremal black hole solutions 
which arise from double-extreme single centre
BPS black hole solutions,  which are supported by constant scalar fields $Y_0^I$, by turning on a non-extremality parameter $\mu$.
These are therefore near-extremal Reissner-Nordstrom
black hole solutions that are supported by constant scalar fields $Y_0^I$,  with line element given by ($\rho > \rho_+ > 0$)
\bea
ds_4^2 = - \frac{ (\rho - \rho_+) (\rho - \rho_-)}{\rho^2} \, dt^2 + \frac{\rho^2}{(\rho - \rho_+) (\rho - \rho_-)} \, d\rho^2 + 
\rho^2 
\left( d \theta^2 + \sin^2 \theta \, d \varphi^2 \right) \;.
\label{lirn}
\eea
Defining the non-extremality parameter $\mu$ and the constant $\Sigma$ in terms of the outer and inner horizons $r_{\pm}$, respectively,
\bea
\mu = \frac12 (\rho_+ - \rho_-) \geq 0 \;\;\;,\;\;\; \Sigma =  \frac12 (\rho_+ + \rho_-) > 0\;,
\eea
and introducing $R = \rho - \Sigma$, the line element \eqref{lirn} can be written as ($R > \mu \geq 0$)
\bea
ds_4^2 = - \frac{R^2 - \mu^2}{ ( R + \Sigma)^2} \, dt^2 + \frac{( R + \Sigma)^2}{R^2 - \mu^2} \, dR^2 + (R + \Sigma)^2 \, 
\left( d \theta^2 + \sin^2 \theta \, d \varphi^2 \right) \;.
\label{demu}
\eea
We recall the relation \cite{Goldstein:2014gta}
\bea
\Sigma^2 = \Sigma_0^2 + \mu^2 \;,
\label{s0s}
\eea
where $ \pi \Sigma_0^2$ denotes the entropy of the extremal Reissner-Nordstrom black hole, which is described by \eqref{demu} with
$\mu=0$, and whose horizon is at $R=0$.

The non-extremal Reissner-Nordstrom
black hole solution possesses temperature 
\bea
T = \frac{\rho_+ - \rho_-}{4 \pi \rho_+^2} = \frac{\mu}{2 \pi (\Sigma + \mu)^2} \;,
\eea
Wald entropy \cite{Iyer:1994ys}
\bea
{\cal S}_{\rm Wald} = \pi  (R + \Sigma)^2 \vert_{R = \mu} = \pi \, (\mu + \Sigma)^2 \;,
\label{entronone}
\eea
and ADM mass
\bea
M = \Sigma \;.
\eea
Working to lowest order in $\mu/\Sigma_0$, these quantities become
\bea
T =  \frac{\mu}{2 \pi \Sigma_0^2} \;\;\;,\;\;\; {\cal S}_{\rm Wald} = \pi \left( \Sigma_0^2 + 2 \mu \Sigma_0 \right) \;\;\;,\;\;\; M = \Sigma_0 + 
\tfrac12 \frac{\mu^2}{\Sigma_0} \;.
\label{tsml}
\eea
These quantities satisfy the first law $ \delta M = T \, \delta S_{\rm Wald} $, where the variation $\delta$ refers to a change
of the non-extremality  parameter $\mu$.

Next, let us consider a different
 solution to the four-dimensional equations of motion, namely ($R > \mu \geq 0$)
\bea
ds_4^2 = - \frac{R^2 - \mu^2}{ \Sigma_0^2} \, dt^2 + \frac{\Sigma_0^2}{R^2 - \mu^2} \, dR^2 + \Sigma_0^2 \, 
\left( d \theta^2 + \sin^2 \theta \, d \varphi^2 \right) \;,
\label{geompro}
\eea
which is the line element of a product metric, describing a two-dimensional black hole and a two-sphere of constant radius $\Sigma_0$.
When $\mu =0$, this product geometry describes the near-horizon geometry of an extremal black hole, which we will take to be BPS.
This two-dimensional black hole is an exact solution of the two-dimensional equations of motion, and it can be put into
the form \eqref{2dBPS}, as follows. First 
note that the two-dimensional black hole has constant curvature $R_2 = 2 /\Sigma_0^2$. Thus, the role of $v_1$ in \eqref{2dBPS} is now
played by $\Sigma_0^2$, i.e. $v_1 = \Sigma_0^2$.
Then, performing the coordinate transformation $dr = \Sigma_0 \, dR / \sqrt{  R^2 - \mu^2} $, i.e.
\bea
e^{r / \Sigma_0} = \frac{\mu}{2} \, \sqrt{ \frac{ R + \sqrt{ R^2 - \mu^2} }{R -  \sqrt{ R^2 - \mu^2} } } =  \frac{ R + \sqrt{ R^2 - \mu^2} }{2} \;,\;
e^{- r / \Sigma_0} =
2 \,  \frac{ R - \sqrt{ R^2 - \mu^2} }{ \mu^2} \;,
\label{coorrR}
\eea
we obtain
\bea
ds_4^2 = - \frac{1}{ \Sigma_0^2} \, \left( e^{r/\Sigma_0} - \frac{\mu^2}{4} e^{-r/\Sigma_0} \right)^2 dt^2 + dr^2 + \Sigma_0^2 \, 
\left( d \theta^2 + \sin^2 \theta \, d \varphi^2 \right) \;,
\label{bhmu}
\eea
which is of the form \eqref{2dBPS} with $\alpha = 1/\Sigma_0$ and $\beta = - \mu^2/4 \Sigma_0 \leq 0$, and with constant $v_2 = \Sigma_0^2 = v_1$.

In the following, we define $h_{tt, 0}$ to be
\bea 
h_{tt, 0} =  - \frac{1}{ \Sigma_0^2} \, \left( e^{r/\Sigma_0} - \frac{\mu^2}{4} e^{-r/\Sigma_0} \right)^2 \;.
\label{htt0}
\eea
The location of the horizon of the two-dimensional black hole is at
$\sqrt{-h_0} \vert_{r_h} = 0$, 
\bea
e^{2 r_h/\Sigma_0} = \frac{\mu^2}{4} \;,
\label{2dbhhor}
\eea
its temperature is
\bea
T = \frac{1}{2 \pi} \, \partial_r \sqrt{-h_0}\vert_{r_h} = \frac{ \mu }{2 \pi \, \Sigma_0^2} \;,
\label{tempbh}
\eea
and its Wald entropy is 
\bea
{\cal S}_{\rm Wald} = \pi \,  v_2 = \pi \, \Sigma_0^2 \;.
\eea
The entropy of the two-dimensional black hole differs from the entropy \eqref{tsml} of the four-dimensional non-extremal 
Reissner-Nordstrom black hole. In the following, we will view the solution \eqref{bhmu} as a background solution
and describe how to regard the non-extremal Reissner-Nordstrom solution as a deformation of the former, to lowest order in the deformation parameter 
$\mu/\Sigma_0$.

First, we note that the metric factors in the line element  \eqref{demu} can be written in terms of three ratios, namely $R/\Sigma_0, \, \mu/\Sigma_0$ and
$\mu/R$, where we recall the relation \eqref{s0s}. Since $\mu/R = (\mu /\Sigma_0) (\Sigma_0 / R)$, we take $R/\Sigma_0$ and $\mu/\Sigma_0$ to be the independent ratios. We multiply $R$ and $\mu$ by a factor $\varepsilon$, thereby attaching a factor  $\varepsilon$ to 
each of these two ratios.  We also rescale $t \rightarrow t/\varepsilon$. Subsequently we expand 
the line element \eqref{demu} to lowest order in $\varepsilon$. At first order in $\varepsilon$, we obtain 
\bea
ds_4^2 &=& - \frac{R^2 - \mu^2}{ \Sigma_0^2} \, \left( 1 - 2  \, \varepsilon \, \frac{R}{\Sigma_0} \right) dt^2 
+ \frac{\Sigma_0^2 ( 1 + 2 \, \varepsilon \, \frac{R}{ \Sigma_0})}{R^2 - \mu^2} \, dR^2  \nonumber\\
&& + \Sigma_0^2 \,
\left( 1 + 2 \, \varepsilon \,  \frac{R}{\Sigma_0} \right)  
\left( d \theta^2 + \sin^2 \theta \, d \varphi^2 \right) \;.
\label{demuapp}
\eea
This describes a first-order deformation of the product
geometry \eqref{geompro}, and hence of \eqref{bhmu}. For consistency with the first-order expansion in $\varepsilon$,
we introduce a cutoff $R_c$  for the radial coordinate, i.e. $R_c \geq R \geq \mu$, and
we demand $\Sigma_0 \gg R_c \gg  \mu$.

Next, 
we
perform the coordinate transformation
\bea
dr = \Sigma_0 \, \frac{ \left( 1 + \varepsilon \, \frac{R}{\Sigma_0} \right) }{\sqrt{ R^2 - \mu^2 }} \, dR \;,
\eea
which results in 
\bea
\frac{r}{\Sigma_0} = \log \frac{ R + \sqrt{ R^2 - \mu^2} }{2} + \varepsilon \, \frac{ \sqrt{R^2 - \mu^2}} {\Sigma_0} =
\log \frac{ R + \sqrt{ R^2 - \mu^2} }{2} + \frac{\varepsilon}{\Sigma_0} \left( e^{r/\Sigma_0} - \frac{\mu^2}{4} e^{-r/\Sigma_0} \right) \;, \nonumber\\
\eea
where, at first order in $\varepsilon$, we have made use of \eqref{coorrR} in the second term.
Thus, to first order in $\varepsilon$, we obtain
\bea
\frac{ R + \sqrt{ R^2 - \mu^2} }{2} &=& 
e^{r / \Sigma_0}  \left( 1 - \frac{\varepsilon}{\Sigma_0} \left( e^{r/\Sigma_0} - \frac{\mu^2}{4} e^{-r/\Sigma_0} \right)
 \right) \;, \nonumber\\
  2 \,  \frac{ R - \sqrt{ R^2 - \mu^2} }{ \mu^2}  &=&
 e^{-r / \Sigma_0} 
 \left( 1 + \frac{\varepsilon}{\Sigma_0} \left( e^{r/\Sigma_0} - \frac{\mu^2}{4} e^{-r/\Sigma_0} \right)
 \right) \;,
\eea
and hence,
\bea
\sqrt{ R^2 - \mu^2}  &=& e^{r / \Sigma_0}  \left( 1 - \frac{\varepsilon}{\Sigma_0} \left( e^{r/\Sigma_0} - \frac{\mu^2}{4} e^{-r/\Sigma_0} \right)
 \right) \nonumber\\
 &&- \frac{\mu^2}{4}  \,  e^{-r / \Sigma_0} \left( 1 + \frac{\varepsilon}{\Sigma_0} \left( e^{r/\Sigma_0} - \frac{\mu^2}{4} e^{-r/\Sigma_0} \right) \right)\;.
\eea
Then, the metric \eqref{demuapp} takes the form \eqref{backgr}
with 
\bea
v_2 &=&  \Sigma_0^2 + \chi \;, \nonumber\\
\chi &=& 2 \, \varepsilon \, \Sigma_0  \left( e^{r/\Sigma_0} + \frac{\mu^2}{4} \, e^{-r/\Sigma_0} \right)   \;,\nonumber\\
h_{tt} &=& - \frac{1}{\Sigma_0^2} \left( e^{r/\Sigma_0} - \frac{\mu^2}{4} \, e^{-r /\Sigma_0} \right)^2
\left( 1 - \frac{4 \, \varepsilon}{\Sigma_0}  
\left(  e^{r/\Sigma_0} + \frac{\mu^2}{4} \, e^{-r /\Sigma_0} \right) \right) \;.
\label{apprdem2}
\eea
Hence, at first order in $\varepsilon$, we obtain
\bea
\sqrt{- h} = \frac{1}{\Sigma_0}  \left( e^{r/\Sigma_0} - \frac{\mu^2}{4} \, e^{-r /\Sigma_0} \right)
\left( 1 - \frac{2 \, \varepsilon}{\Sigma_0}  
\left(  e^{r/\Sigma_0} + \frac{\mu^2}{4} \, e^{-r /\Sigma_0} \right) \right) \;.
\eea
Writing $\sqrt{- h} = \sqrt{- h_0} + \sqrt{- h_1}$, where $h_{tt,0}$ is given in \eqref{htt0},
we infer  
\bea
\sqrt{- h_1} = 
- \frac{\varepsilon}{\Sigma_0}  \left( e^{r/\Sigma_0} - \frac{\mu^2}{4} \, e^{-r /\Sigma_0} \right)
\frac{2}{\Sigma_0}  
\left(  e^{r/\Sigma_0} + \frac{\mu^2}{4} \, e^{-r /\Sigma_0} \right) = - \frac{\sqrt{-h_0}}{\Sigma_0^2} \, \chi \;.
\eea
This is of the form \eqref{pertsol}, 
  with 
\bea
\alpha = \frac{1}{\Sigma_0} \;\;\;,\;\;\; \beta = -  \frac{\mu^2}{4 \Sigma_0} \;\;\;,\;\;\; \nu = 2 \, \varepsilon \, \Sigma_0 \;\;\;,\;\;\; \vartheta = 
\varepsilon \, \frac{ \Sigma_0 \mu^2}{2} \;.
\label{albe}
\eea
These constant values satisfy \eqref{avbn2}, as they must.

We note that if we define the constant
\bea
c_0 \equiv (\varepsilon \, \Sigma_0 \mu) ^2 \;,
\label{corel}
\eea
we obtain the relations
\bea 
\vartheta = \frac{c_0}{\nu} \;\;\;,\;\;\; \beta = - \frac{\alpha}{\nu^2} \, c_0 \;.
\label{bvc0}
\eea
In  section \ref{sec:time-dep} we will consider time-dependent perturbations of the non-extremal Reissner-Nordstrom black hole.
These are governed by \eqref{pertsol2}. We will show that when switching off the time-dependent perturbation, we obtain 
\eqref{bvc0}.

Finally, let us return to the condition \eqref{condnu} that the perturbation $\chi$ must satisfy.
At large $r$, the metric factor $h_{tt} $ behaves as
\bea
h_{tt} = 
- \left( \frac{e^{r /\Sigma_0} }{\Sigma_0} \right)^2 \left( 1 - 4 \varepsilon  \frac{e^{ r /\Sigma_0} }{\Sigma_0} \right) \;.
\eea
To ensure that the term proportional to $\varepsilon$ is suppressed \cite{Castro:2018ffi}, we cut off the radial direction
at $r_c < + \infty$, and we demand 
\bea
  \frac{e^{ r_c /\Sigma_0} }{\Sigma_0} \ll 1 \;,
  \label{condcut}
\eea
which is equivalent to the condition $R_c \ll \Sigma_0$ discussed above.

In the coordinates used in \eqref{demuapp}, the horizon is still at $R = \mu$, and 
the first-order deformed solution \eqref{demuapp} exhibits the following change in the entropy when compared with
the entropy of the exact solution \eqref{geompro},
\bea
\Delta {\cal S} = \varepsilon \,  2 \pi \, \mu \, \Sigma_0 \;.
\label{chaent}
\eea
To lowest order in $\varepsilon$, the temperature of the deformed solution, in the coordinates used in \eqref{demuapp}, is given by $T = \mu/(2 \pi \Sigma_0^2)$, and thus agrees with \eqref{tempbh}. Hence, to lowest order in $\varepsilon$,
 the deformed solution \eqref{demuapp} 
exhibits the change 
\bea
T \Delta {\cal S} =\varepsilon \,  \frac{\mu^2}{\Sigma_0} 
\eea
when compared with \eqref{geompro}. Since $T \Delta {\cal S}$ is non-vanishing, the deformed solution should carry mass,
as required by the first law of black hole mechanics.  To be able to infer the mass of the deformed solution, we
observe that working to second
order in $\varepsilon$, the four-dimensional
asymptotically flat Reissner-Nordstrom black hole can be viewed as a 
two-dimensional $\varepsilon$-deformed background.
In order to see this, 
in \eqref{demu} we perform 
the rescaling $\mu \rightarrow  \varepsilon \, \mu$, $R \rightarrow \varepsilon \, R$ and $t \rightarrow t/\varepsilon$, so 
as to be working with the same coordinates used in \eqref{demuapp}. 
We also expand $\Sigma$ to second order in $\varepsilon$.
Then, \eqref{demu} becomes
\bea
ds_4^2 = - \frac{R^2 - \mu^2}{ ( \varepsilon R + \Sigma)^2} \, dt^2 + \frac{( \varepsilon R + \Sigma)^2}{R^2 - \mu^2} \, dR^2 + (\varepsilon R + \Sigma)^2 \, 
\left( d \theta^2 + \sin^2 \theta \, d \varphi^2 \right) \;,
\label{demu2}
\eea
with $\Sigma$ replaced by $\Sigma_0 + \ft12 \varepsilon^2 \, \mu^2 / \Sigma_0$. Expanding \eqref{demu2} to second order in $\varepsilon$ gives
 \eqref{demuapp}, with additional terms proportional to  $\varepsilon^2$, which we do not exhibit. On the other hand, when taking the limit $R \gg 1 $ in \eqref{demu2},
 the term proportional to $dt^2$ becomes (up to ${\cal O} (1/R^2$))
 \bea
 -  \left( 1 - \frac{2 }{\varepsilon \, R} \, \left(\Sigma_0 + \ft12 \varepsilon^2 \, \frac{\mu^2 }{\Sigma_0} \right)  \right) \, \frac{dt^2}{\varepsilon^2} \;.
 \eea
 We now identify the coefficient of the term $1/R$ with twice the mass $M$, which we write as $M = M_0 + \Delta M$ with
 \bea
 M_0 = \frac{\Sigma_0}{\varepsilon} \;\;\;,\;\;\; \Delta M =  \ft12 \varepsilon \, \frac{\mu^2}{ \Sigma_0 } \;. 
 \label{mDm}
 \eea
 Note that $M_0$ is independent of $\mu$. 
Then, varying \eqref{mDm} with respect to $\mu$, we establish the validity of the first law for the deformed solution at first
 order in $\varepsilon$,
 \bea
 \delta \Delta M = T \, \delta \Delta {\cal S } \;.
 \eea
  Thus, the above definition of mass is consistent with the first law of black hole mechanics.

 As shown in  \cite{Goldstein:2014gta}, there is a transformation, called Harrison transformation, that maps the exact solution \eqref{geompro}
 to the exact solution \eqref{demu2}. This transformation changes the asymptotic behaviour of the solution, while keeping the horizon fixed at $R=\mu$.
 It acts as follows on the exact solution \eqref{geompro}. If we write the latter as
 \bea
 ds_4^2 = - a^2(R) \, dt^2 + \frac{dR^2}{a^2(R)} + b^2(R) \, d \Omega^2 \;,
 \eea
 then the Harrison transformation acts as \cite{Goldstein:2014gta}
 \bea
 a \rightarrow \Lambda (R) \, a \;\;\;,\;\;\; b \rightarrow \frac{b}{\Lambda (R)} \;.
 \eea
 For the choice
 \bea
\Lambda (R) = \frac{\Sigma_0}{ \varepsilon R + \Sigma } \;,
\eea
the solution \eqref{geompro} gets mapped to the solution \eqref{demu2} with the rescaling $t \rightarrow t/\varepsilon$. 
 To first order in $\varepsilon$, this Harrison transformation precisely yields \eqref{demuapp}.

\subsubsection{Time-dependent perturbations of near-extremal Reissner-Nordstrom solutions \label{sec:time-dep}}

Next, let us consider time dependent perturbations of the near-extremal Reissner-Nordstrom solution.
To do so, we add a time-dependent perturbation to $\chi$ given in \eqref{apprdem2},
\bea
\chi  &=& \nu(t) \,  e^{r/\Sigma_0} + \vartheta (t) \,  e^{-r/\Sigma_0} \nonumber\\
&=&\varepsilon \left( 2 \Sigma_0 + N(t) \right)   e^{r/\Sigma_0} + \varepsilon \left( \frac{\Sigma_0 \, \mu^2}{2 } + \mathbb{T} (t)  \right)  e^{-r/\Sigma_0} \;,
\label{nvt}
\eea
with $N' \neq 0$.
We work again at first order in $\varepsilon$.

The time dependent perturbation $\nu(t)$ will induce a time-dependent deformation of $\beta$, c.f. 
\eqref{pertsol2}. Hence, 
the space-time metric $\sqrt{-h_0}$ (c.f. \eqref{htt0}) is deformed to 
\bea
\sqrt{- h_0} = \frac{1}{\Sigma_0}  \, e^{r/\Sigma_0} + \beta(t) \, e^{-r /\Sigma_0} \;,
\eea
with $\beta (t)$ determined in terms of $\nu(t)$ by the last equation of \eqref{pertsol2}. We write
\bea
\beta(t) = - \frac{\mu^2}{4 \Sigma_0} +   l(t) \;,
\label{lfun}
\eea
where below $l(t)$ will be determined in terms of $N(t)$. Note  that the last equation of \eqref{pertsol2} implies
that $\beta(t)$ is of order $\varepsilon^0$.
According to \eqref{pertsol},
 $\sqrt{-h_0}$ is further deformed into
 $\sqrt{-h} = \sqrt{- h_0}  + \sqrt{- h_1}$, with
\bea
\sqrt{-h_1} = -\frac{\sqrt{-h_0}}{\Sigma_0^2} \, \chi - 2\,  \varepsilon \, \Sigma_0 \,  N'' (t) \;,
\eea
where in this expression we only retain terms up to order $\varepsilon$.

Now recall from \eqref{pertsol2} that $\beta$ and $\vartheta$ are determined in terms of $\nu$.
Using the first relation in  \eqref{pertsol2},
we obtain to first order in $\varepsilon$,
\bea
\vartheta = \frac{\varepsilon}{2 \Sigma_0 + N} \left( \left( \Sigma_0 \, \mu \right)^2 - \frac{\Sigma_0^4}{4} \left( N' \right)^2 \right) \;,
\eea
where we used \eqref{corel}. Then, comparing with \eqref{nvt}, we infer
\bea
\mathbb{T} (t) =- \left(  \frac{2 \Sigma_0 \mu^2 N + \Sigma_0^4 \left( N' \right)^2 }{4 ( 2 \Sigma_0 +N )} \right) \;.
\eea
Using the last relation  in \eqref{pertsol2}, we obtain, using the value for $\alpha$ given in \eqref{albe},
\bea
\beta = - \frac{1}{\Sigma_0 (2 \Sigma_0 + N )^2} \left( (\Sigma_0 \mu)^2 - \frac{\Sigma_0^4}{4} \left( N' \right)^2 \right) - \frac{\Sigma_0^3}{2 (2 \Sigma_0 +N)}
 \, N'' \;.
\eea
Comparing with \eqref{lfun} we infer
\bea
l(t) = \frac{\mu^2}{4 \Sigma_0} \, N \, \frac{(4 \Sigma_0 + N )}{(2 \Sigma_0 + N )^2}  + \frac{\Sigma_0^3}{4 (2 \Sigma_0 + N )^2} \left( N' \right)^2 
 - \frac{\Sigma_0^3}{2 (2 \Sigma_0 +N)} \, N'' \;.
\eea

Next, we will assume the  hierarchy $|N| \ll \mu \ll \Sigma_0$. Then we obtain the expressions 
\bea
\beta (t) &=&  - \frac{\mu^2}{4 \Sigma_0} -   \frac{\Sigma_0^2}{4} \, N''  \;, \nonumber\\
\nu (t) &=& \varepsilon \left( 2 \Sigma_0 + N(t) \right) \;\;\;,\;\;\; \vartheta = \varepsilon \,  \frac{ \mu^2 \left( 2 \Sigma_0 - N(t)\right) - \ft12 \Sigma^3_0 (N')^2 }{4 }  \;.
\label{rel1stf}
\eea
When switching off $N(t)$, these expressions reduce to \eqref{bvc0}.

In section \ref{sec:calogero} we will discuss the boundary action that describes the dynamics of $\nu (t)$. This boundary action 
depends on the constant $c_0$, c.f.  \eqref{dffac}. If we take the background to be the near-extremal Reissner-Nordstrom black hole,
this parameter takes the value \eqref{corel}.

\section{Variational principle and holographic renormalization \label{sec:calogero}}

The equations of motion for the dynamical fields 
$h_{ij}, e^I, f_J, v_2, Y^I$ are obtained by subjecting the action to a variational principle with a prescribed set of boundary conditions
that are imposed at the boundaries of the two-dimensional space-time manifold $M$. Here, for simplicity, we will consider just one boundary 
$\partial M$, which we take to be time-like. In view of the condition  \eqref{condnu}, we take 
$\partial M$ to be 
the timelike line at $r = r_c \gg 1 $. We will demand that the expressions given below are finite when removing the cutoff, i.e. when $r_c 
\rightarrow + \infty$. Hence, they will be independent of the cutoff at leading order, and they will coincide 
with the expressions computed in the limit $r \rightarrow + \infty$, up to exponentially suppressed cutoff dependent terms. 
Hence, in the following, we will take the boundary
$\partial M$ to be at $r = + \infty$ and ignore exponentially suppressed cutoff dependent terms,  for simplicity.
Obtaining a consistent variational principle for the dynamical fields requires adding suitable boundary terms
to the bulk action. Let us consider the term
\bea
- \frac{1}{2 G_4} \,\sqrt{-h} \, v_2 \, R_2 
\label{R2}
\eea
in the bulk Lagrangian \eqref{lagHef}. A consistent variational principle for the space-time metric $h_{ij}$ requires adding a 
Gibbons-Hawking boundary term of the form 
\bea
\frac{1}{G_4} \, \int_{\partial M}  dt \, \sqrt{-\gamma} \, v_2 \, K \;,
\label{K}
\eea
where $K$ denotes the trace of the extrinsic curvature tensor, and $\sqrt{-\gamma}$ denotes the induced metric on $\partial M$.  Then, varying the combination
\eqref{R2} and \eqref{K} with respect to the space-time metric results in
\bea
\frac{1}{2 G_4} \, \int_{\partial M}  dt \, \sqrt{-\gamma} \, \partial_r v_2 \, \gamma^{tt} \, \delta \gamma_{tt} \;,
\label{varRK}
\eea
a result that will be used shortly. 
In a situation where $\gamma_{tt}$ diverges at the boundary $\partial M$ and hence cannot be kept fixed at the boundary,
this term will have to be cancelled by the variation of additional boundary terms \cite{Papadimitriou:2010as,Cvetic:2016eiv}. 
Proceeding in this manner, one obtains a combined bulk-boundary action that implements a consistent variational principle with 
appropriately prescribed boundary 
conditions.

In the following, we discuss the renormalized on-shell action $S_{ren}$ \cite{Skenderis:2002wp}
 for the perturbed solution  \eqref{pertsol}. We restrict ourselves to the case when ${\cal Y}^I =0$.
 The resulting renormalized action has been derived before in the context of 
 five-dimensional pure Einstein gravity with cosmological constant reduced to two dimensions \cite{Castro:2018ffi}, and three-dimensional topologically massive gravity reduced
to two dimensions \cite{Castro:2019vog}.
Here we follow \cite{Castro:2019vog}.

The renormalized on-shell action is a functional of the modes $\alpha(t), \nu(t)$ and
$\mu^I(t), {\tilde \mu}_ I (t)$ (c.f. \eqref{AtAr}) that parametrize the perturbed solution \eqref{fopert}, c.f. 
\eqref{2dBPS},  \eqref{pertsol} and 
 \eqref{pertsol2} (recall that $\beta(t)$ and $\vartheta(t)$ are
expressed in terms of $\alpha(t)$ and $\nu(t)$). Then, if one varies $S_{ren}$ with respect to $\alpha(t), \nu(t), \mu^I(t), {\tilde \mu}_ I (t)$, 
one is considering variations in the space of solutions to the field equations, 
\bea
\delta S_{ren} = \int_{\partial M} \, dt \, 
 \alpha 
\left( \hat{\pi}^{tt} \, \delta (- \alpha^2)  + \hat{\pi}_{v_2} \, \delta \nu + \hat{\pi}_I \, \delta \mu^I + \hat{\tilde \pi}_I \,
\delta {\tilde \mu}^I  
\right) \;,
\label{varSren}
\eea
where the $\hat{\pi}$ are evaluated on-shell and finite. In the boundary theory, 
$\alpha(t), \nu(t), \mu^I(t), {\tilde \mu}_ I (t)$ act as sources.
The variations $ \delta (- \alpha^2),  \delta \nu, \delta \mu^I , \delta {\tilde \mu}^I $ 
implement a variational principle with the following
boundary conditions on the fields,
\bea
\gamma_{tt} &=& - \alpha^2 \, e^{2r/\sqrt{v_1}}\;, \nonumber\\
\chi &=& \nu \, e^{r/\sqrt{v_1}} \;, \nonumber\\
A_t^{ren \, I} &=&   \mu^I \;\;\;,\;\;\;  {\tilde A}_{I t}^{ren} =  {\tilde \mu}_I \;,
\eea
subject to
\bea
\delta \gamma_{tt} &=& \delta (- \alpha^2) \, e^{2r/\sqrt{v_1}}\;, \nonumber\\
\delta v_2 &=& \delta \chi =  \delta \nu \, e^{r/\sqrt{v_1}} \;, \nonumber\\
\delta A_t^{ren \, I} &=&  \delta \mu^I \;\;\;,\;\;\; \delta {\tilde A}_{I t}^{ren} = \delta {\tilde \mu}_I \;.
\label{sourfie}
\eea
Here, $ A_t^{ren \, I}$ and ${\tilde A}_{I t}^{ren} $ are fields that differ from $A_t^I$ and ${\tilde A}_{I t}$ and that are such
that they  approach $\mu^I $ and ${\tilde \mu}_I $ at $r  = + \infty$, respectively. We refer to \cite{Cvetic:2016eiv,Aniceto:2020saj} for the construction
of these fields.
Then, \eqref{varSren} can also be written as 
\bea
\label{renacvar}
\delta S_{ren} = \int_{\partial M} dt  \left( \pi^{tt} \, \delta \gamma_{tt} + \pi_{v_2} \, \delta v_2 + \pi_I \, \delta A_t^{ren \, I} + 
{\tilde \pi}^I \, \delta {\tilde A} _{I t}^{ren } 
\right) \;, 
\label{varsren}
\eea
where the $\pi$ are evaluated on-shell and finite,
\bea
\hat{ \pi}^{tt} = 
\lim_{r \rightarrow + \infty} \frac{  e^{2r/\sqrt{v_1}}}{\alpha} 
\pi^{tt}  \;\;\;,\;\;\;
 \hat{\pi}_{v_2}  = \lim_{r \rightarrow +  \infty}  \frac{  e^{r/\sqrt{v_1}}}{\alpha}  \pi_{v_2} \;\;\;,\;\;\; 
\hat{\pi}_I = \frac{\pi_I}{\alpha} \;\;\;,\;\;\,  
\hat{{\tilde \pi}}^I = \frac{ {\tilde \pi}^I}{\alpha} \;. 
\label{hpp}
\eea

Following \cite{Castro:2019vog}, we now
determine $S_{ren}$ to first order in $\epsilon$ (c.f. section \ref{sec:flucana}), by
 first obtaining the $\hat{\pi}$ and subsequently
integrating these expressions. Thus, we take the terms in \eqref{varSren}  to be 
of order $\epsilon$. 
This implies that 
$\hat{\pi}^{tt}$ is of order $\epsilon$, while  $\hat{\pi}_{v_2}$ is of order $\epsilon^0$.

Varying \eqref{K} with respect to $v_2$, i.e. $\delta v_2 = \delta \chi$ with $\delta \chi$ given in \eqref{sourfie}, yields
\bea
\frac{1}{G_4} \, \int_{\partial M}  dt \, \sqrt{-\gamma} \, \delta \chi \, K \;. 
\label{Kdch}
\eea
In this expression, since $\delta \chi$ is of order $\epsilon$, $K$ is evaluated in the background metric $\sqrt{-h_0}$,
\bea
K = \frac{1}{\sqrt{v_1}} \left[ 1 - 2\, \frac{\beta}{\alpha} \, e^{-2 r /\sqrt{v_1}} + {\cal O} \left( e^{-4r/\sqrt{v_1}} \right)  \right] \;,
\label{Kasym}
\eea
and hence \eqref{Kdch}
gives rise to a divergence $e^{2r/\sqrt{v_1}}$. To cancel this divergence, we add the following boundary term to \eqref{K}  \cite{Castro:2019vog},
\bea
\frac{1}{G_4} \, \int_{\partial M}  dt \, \sqrt{-\gamma} \, \left( v_2 \, K - \frac{\chi}{\sqrt{v_1}} \right) \;.
\label{Kimp}
\eea
Varying this expression with respect to $\chi$ yields
\bea
\frac{1}{G_4} \, \int_{\partial M}  dt \, \sqrt{-\gamma} \, \delta \chi \left( K - \frac{1}{\sqrt{v_1}} \right) \;,
\label{Kimpvar}
\eea
which is finite.  On the other hand, varying the second term in \eqref{Kimp} with respect to $\gamma_{tt}$ gives
\bea
- \frac{1}{2 G_4} \, \int_{\partial M}  dt \, \sqrt{-\gamma} \, \frac{\chi}{\sqrt{v_1}}   \, \gamma^{tt} \, \delta \gamma_{tt} \;.
\eea
Adding this to \eqref{varRK} gives
\bea
\frac{1}{2 G_4} \, \int_{\partial M}  dt \, \sqrt{-\gamma} \, \left( \partial_r v_2  -   \frac{\chi}{\sqrt{v_1}}  \right) \, \gamma^{tt} \, \delta \gamma_{tt} \;,
\label{stvar}
\eea
which is finite, as can be verified by using \eqref{pertsol}.  Thus, the variation of the sum of \eqref{R2} and \eqref{Kimp} 
with respect to $\chi$ and with respect to the space-time metric yields a finite result at order $\epsilon$. 
Using \eqref{hpp} we infer
\bea
\hat{\pi}^{tt} = \frac{1}{2 G_4}  \lim_{r \rightarrow + \infty} \left[\frac{
\sqrt{-\gamma} \, \gamma^{tt} \,  e^{2 r /\sqrt{v_1}} }{\alpha} \left( \partial_r v_2  -   \frac{\chi}{\sqrt{v_1}}  \right) \right]= \frac{\vartheta}{G_4 \, \sqrt{v_1} \, \alpha^2} \;,
\label{relp1}
\eea
which is of order $\epsilon$.
Next, using \eqref{Kimpvar} and  \eqref{hpp} we obtain
\bea
\hat{\pi}_{v_2} = \frac{1}{G_4}  \lim_{r \rightarrow + \infty} \left[
\frac{\sqrt{-\gamma} \,  e^{ r /\sqrt{v_1}} }{\alpha} \left( K -   \frac{1}{\sqrt{v_1}}  \right) \right]= - \frac{2}{G_4 \, \sqrt{v_1}} \frac{\beta}{\alpha} \;,
\label{relp2}
\eea
which is of order $\epsilon^0$. This expression will receive a correction of order $\epsilon^1$ induced by the variation of the
term $(\partial v_2)^2/v_2$
in the bulk Lagrangian \eqref{lagHef}, but since here we only determine $S_{ren}$ to order $\epsilon$, we drop this correction term.
Using \eqref{avbnmod}, we note the relation \cite{Castro:2018ffi}
\bea
 \frac{\alpha^2 \, {\hat \pi}^{tt}}{\nu} +  \frac{1}{G_4 \sqrt{v_1}} \, \frac{v_1}{2 \alpha \, \nu} 
\left( \frac{\nu'}{\alpha} \right)' 
  = \frac{{\hat \pi}_{v_2} }{2 } \;.
\label{relpp}
\eea

Next, using the expression \eqref{pertsol2} and integrating
\bea
\alpha \, \hat{\pi}^{tt} = - \frac{\partial S_{ren}}{\partial (\alpha^2)}  = - \frac{1}{2 \alpha} \frac{\partial S_{ren}}{\partial \alpha}
 \;\;\;,\;\;\; \alpha \,  \hat{\pi}_{v_2} = \frac{ S_{ren}}{\partial \nu}
\eea
with respect to $\alpha$ and $\nu$ we obtain \cite{Castro:2018ffi,Castro:2019vog}, up to a total derivative term, and up to
terms proportional to $\mu^I$ and ${\tilde \mu}_I$,
\bea
S_{ren} = -  \int_{\partial M} dt \, \frac{2}{G_4 \, \sqrt{v_1} } \left( c_0 \, \frac{\alpha}{\nu} + \frac{v_1}{4} \frac{ \left( \nu' \right)^2}{\alpha \, \nu} \right) \;.
\label{ross}
\eea

Finally, using that on-shell \cite{Aniceto:2020saj},
\bea
\alpha \, \hat{\pi}_I  =   - q^I   = \frac{\partial S_{ren}}{\partial \mu^I}   \;\;\;,\;\;\; \alpha \, \hat{\tilde \pi}_I = p_I  = \frac{\partial S_{ren}}{\partial {\tilde \mu}_I}  \;,
\eea
and integrating with respect to $ \mu^I, {\tilde \mu}_I$, we obtain for the renormalized on-shell action to first order in $\epsilon$,
\bea
S_{ren} =   \int_{\partial M} \, dt \left( - 
\frac{2}{G_4 \, \sqrt{v_1} } \left( c_0 \, \frac{\alpha}{\nu} + \frac{v_1}{4} \frac{ \left( \nu' \right)^2}{\alpha \, \nu} \right)  -q_I \, \mu^I + p^I \,  {\tilde \mu}_I
\right) \;.
\label{ross2}
\eea
This is the renormalized action based on the perturbed solution \eqref{pertsol} for the case ${\cal Y}^I =0$, and assuming $\nu' \neq 0$. 
Substituting $\nu (t) = x^2(t)$ in \eqref{ross}, we note that the renormalized on-shell action \eqref{ross} is 
related to the DFF action \cite{deAlfaro:1976vlx},
\bea
S_{\rm DFF} = -  \int_{\partial M} dt \, \frac{2}{G_4 \, \sqrt{v_1} } \left( c_0 \, \frac{\alpha(t) }{x^2(t)} + 
\frac{v_1}{\alpha(t)}  (x')^2 \right) \;.
\label{dffac}
\eea
In this action, the special geometry data of the four-dimensional theory is 
encoded in  $v_1 = v_{2,0} = e^{-{ \cal K}(Y_0)} G_4$.

\section{ Asymptotic symmetries}

The AdS/CFT correspondence posits that a perturbation of bulk parameters, such as black hole mass, must be holographically encoded in
the boundary CFT. 
Using the anomalous transformation behaviour of $\hat{\pi}^{tt}$ and $\hat{\pi}_{v_2}$ under asymptotic symmetry transformations 
\cite{Cvetic:2016eiv,Castro:2018ffi}, we show how to holographically encode the change of black hole mass under perturbations, i.e. $\Delta M$, as well
as the hot attractor invariant \cite{Goldstein:2014gta} associated to it, i.e. $v_{2,0}$.
We use the notation of \cite{Aniceto:2020saj}.

We consider the following residual transformations that leave the Fefferman-Graham form of the metric \eqref{backgr}  and the gauge \eqref{rgA}
of the gauge fields invariant \cite{Cvetic:2016eiv},
\bea
\delta \alpha (t)  &=& \partial_t \left(( \varepsilon (t) \, \alpha (t) \right) + \frac{\sigma (t) }{\sqrt{v_1}} \, \alpha (t) \;, \nonumber\\
\delta \nu  (t)&=& \varepsilon (t) \, \partial_t \nu (t) +  \frac{\sigma (t) }{\sqrt{v_1}} \, \nu (t) \;.
\label{rest}
\eea
These residual transformations depend on two arbitrary functions\footnote{The function $\varepsilon (t)$ used in this section should not be confused with the expansion parameter $\varepsilon$ used in the previous sections.} of $t$, $\varepsilon (t)$ and $\sigma (t)$. 
Using the expression for $\beta$ given in \eqref{pertsol2}, one verifies that the residual transformations \eqref{rest}  correctly
induce the transformation law of $\beta$ \cite{Cvetic:2016eiv},
\bea
\delta \beta =  \partial_t \left(\varepsilon \, \beta \right)- \frac{\sigma}{\sqrt{v_1}} \, \beta 
 - \frac{\sqrt{v_1}}{2} \, \partial_t \left( \frac{\partial_t \sigma}{\alpha} \right) \;.
 \label{delbe}
 \eea
Then,  $\hat{\pi}^{tt}$ and $\hat{\pi}_{v_2}$ given in \eqref{relp1} and in \eqref{relp2}, respectively,
transform as 
\bea
\delta \left( \alpha^2 \,  \hat{\pi}^{tt}\right)  &=&   \frac{1}{G_4 \, \sqrt{v_1} } \, \delta \vartheta
=
 \varepsilon \, \partial_t \left( \alpha^2 \,  \hat{\pi}^{tt}\right )- \frac{\sigma}{\sqrt{v_1}} \, \left( \alpha^2 \,  \hat{\pi}^{tt}\right)
- \frac{1}{2 G_4} \,
\frac{\partial_t \nu \partial_t \sigma}{\alpha^2} \;, \\
\delta \hat{\pi}_{v_2}  &=& - \frac{2}{G_4 \, \sqrt{v_1} } \, \delta \left( \frac{\beta}{\alpha} \right)  = 
 \varepsilon \, \partial_t   \hat{\pi}_{v_2}  
  - 2 \frac{\sigma}{\sqrt{v_1}}  \,  \hat{\pi}_{v_2}    + \frac{1 }{\alpha \, G_4}  \partial_t\left( \frac{\partial_t \sigma}{\alpha}  \right) \,.
  \nonumber 
\eea
Now we consider asymptotic symmetries, that is residual transformations that leave the sources invariant, i.e. $\delta \alpha = \delta \nu = 0$.
The condition $\delta \alpha = 0$ results in \cite{Cvetic:2016eiv}
\bea
\sigma = - \sqrt{v_1}  \partial_+ \zeta \;\;\;,\;\;\; \zeta = \varepsilon \, \alpha \;\;\;,\;\;\; \partial_+ = \frac{1}{\alpha} \, \partial_t \;,
\label{sigze}
\eea
while the condition $\delta \nu =0$ results in \cite{Castro:2018ffi}
\bea
\zeta \, \partial_+ \nu = \nu \, \partial_+ \zeta \;,
\eea
which is solved by
\bea
\zeta = k \, \nu \;\;\;,\;\;\; k > 0 \;.
\label{zetnu}
\eea
Then we get
\bea
\delta \left( \alpha^2 \, \hat{\pi}^{tt} \right)  &=& 
  \zeta\, \partial_+  \left( \alpha^2 \, \hat{\pi}^{tt} \right) +( \partial_+ \zeta) \,  \left( \alpha^2 \, \hat{\pi}^{tt} \right) + \frac{\sqrt{v_1}}{2 G_4} \,
\frac{(\partial_+^2 \zeta) \partial_+ \zeta}{k} \;, \nn\\
\delta \hat{\pi}_{v_2} &=&  \zeta\, \partial_+  \hat{\pi}_{v_2} + 
2 \left( \partial_+ \zeta \right) \hat{\pi}_{v_2}  -  
\frac{\sqrt{v_1}}{G_4} \,
 \partial_+^3 \zeta \,.
\eea
Note that the transformation law of $\alpha^2 \,  \hat{\pi}^{tt}$ is not of the standard type: its scaling weight 
 (the coefficient of the term $( \partial_+ \zeta) \,  \left( \alpha^2 \, \hat{\pi}^{tt} \right) $)
is $\Delta =1$, not $\Delta =2$. A combination that involves  $\alpha^2 \,  \hat{\pi}^{tt}$  and has scaling weight $\Delta =2$ is the one given in \eqref{relpp}:
since it equals  $\tfrac12 \hat{\pi}_{v_2}$, it transforms 
in the same way as $\hat{\pi}_{v_2}$ 
under asymptotic symmetry transformations.

The 
Noether charge associated with the asymptotic symmetry transformation \eqref{zetnu} is the constant $c_0$ \cite{Castro:2018ffi} which,
using \eqref{pertsol2}, can be expressed as 
\bea                                                                                                                                                                                                                                                                                                                               
c_0 = G_4 \sqrt{v_1} \,  \nu \left( \alpha^2 \, \hat{\pi}^{tt} \right) + \frac{v_1}{4} \left( \frac{ \nu'}{\alpha} \right)^2 \;.
\label{copittrel}
\eea
In the specific case of the near-extremal Reissner-Nordstrom black hole, which satisfies \eqref{albe},
the relation \eqref{copittrel} yields
\bea
2 G_4 \sqrt{v_1} \, \alpha^2 \left( \alpha^2 \,  \hat{\pi}^{tt} \right) = \frac{\alpha^3 c_0}{\varepsilon} = 2 \Delta M \;,
\eea
with $\Delta M$ given in \eqref{mDm}. Defining
\bea
 \hat{\pi}^t_t = - \alpha^2 \,  \hat{\pi}^{tt} \;,
 \eea
 and using $\alpha^2=1/v_1$,  
 this becomes
 \bea
 - \frac{G_4 }{\sqrt{v_1} }\, \hat{\pi}^t_t =  \Delta M \;.
\eea

Next, let us consider
a particular rescaling of $\hat{\pi}_{v_2}$, with the intend of relating the rescaled quantity, $\hat{\pi}_{\psi}$, to Wald's entropy
of the BPS solution \eqref{2dBPS}. Namely,
inspection of \eqref{rescalXA} shows that there are two possibilities to perform the rescaling $v_2 \rightarrow \lambda \, v_2$
keeping $G_4$ fixed: either $w$ also scales, or $w$ does not scale.
In the latter case one infers that $(Y^I , \Upsilon) \rightarrow (\lambda \, Y^I, \lambda^2 \Upsilon)$. This is the scaling that was used in \cite{Aniceto:2020saj}
to relate $\hat{\pi}_{\psi}$ to Wald's entropy in the presence of $R_2^2$ interactions. Here we consider the former possibility: we take $w$ to scale
as $w \rightarrow \lambda^{-1} \, w$, in which case we obtain $(Y^I, \Upsilon) \rightarrow (Y^I, \lambda^{-2} \, \Upsilon)$, so that under an
infinitesimal variation we get $\delta v_2 = v_2 \, \delta \lambda, \, \delta Y^I = 0, \, \delta \Upsilon = - 2 \Upsilon \, \delta \lambda$.
This is consistent with the fluctuation analysis at the two-derivative level given in section \ref{sec:flucana}: inspection of \eqref{YdY} and \eqref{pdeY}
shows that we may take $\delta Y^I = {\cal Y}^I = 0$ (we recall that we used \eqref{vupsky} to remove the dependence on $\Upsilon$ in the
two-derivative Lagrangian). Setting 
\bea
v_2 = e^{- \psi} \, G_4 \;,
\eea
and taking $\delta \lambda = - \delta \psi$, we obtain
\bea
\delta v_2 = - \delta \psi \, v_2 \;\;\;,\;\;\; \, \delta Y^I = 0 \;.
\eea
Subsequently, we define
\bea
\hat{\pi}_{\psi} = - v_{2,0} \, \hat{\pi}_{v_2}  = \frac{2\, v_{2,0}}{G_4 \, \sqrt{v_1}} \, \frac{\beta}{\alpha} =\frac{2}{ \sqrt{v_1}} \, 
 \frac{{\cal S}_{\rm Wald}^{\rm extr}} {\pi}  \, \frac{\beta}{\alpha} \;,
\eea
where ${\cal S}_{\rm Wald}^{\rm extr} = \pi v_{2,0} /G_4$ denotes the
background value given in \eqref{2dBPS}, which is the hot attractor invariant for a near-extremal four-dimensional
black hole \cite{Goldstein:2014gta}.
 $\hat{\pi}_{\psi}$ has the anomalous transformation behaviour
\bea
\delta  \hat{\pi}_{\psi} &=&  \zeta\, \partial_+  \hat{\pi}_{\psi}  + 
(2 \partial_+ \zeta )\,  \hat{\pi}_{v_2}  +
\sqrt{{v}_1 }\, \frac{{\cal S}_{\rm Wald}^{\rm extr}}{\pi}
\, \partial_+^3 \zeta \;.
\label{anomppsi}
\eea
Taking 
\bea
\delta \psi = \Xi \, e^{r/\sqrt{v_1}} \;,
\eea
where $\Xi(t)$ acts as a source for an irrelevant operator of scaling dimension $2$ in the boundary theory, the 
variation of the renormalized on-shell action \eqref{varSren}
becomes 
\bea
\delta S_{ren} = \int_{\partial M} \, dt \, 
 \alpha 
\left( \hat{\pi}^{tt} \, \delta (- \alpha^2)  + \hat{\pi}_{\psi} \, \delta \Xi + \hat{\pi}_I \, \delta \mu^I + \hat{\tilde \pi}_I \,
\delta {\tilde \mu}^I  
\right) \;.
\label{varSren2}
\eea

Note that the expression for $\hat{\pi}_{v_2}$ given in \eqref{relp2} is of order $\epsilon^0$. It would be interesting to verify
whether at order $\epsilon^1$ $\hat{\pi}_{\psi} $ also captures the change in the entropy \eqref{chaent} of the near-extremal Reissner-Nordstrom
black hole.

\vskip 5mm

\subsection*{Acknowledgements}
This work was partially
supported by FCT/Portugal through 
CAMGSD, IST-ID,
projects UIDB/04459/2020 and UIDP/04459/2020,
through the LisMath PhD fellowship PD/BD/ 128415/2017 (P. Aniceto) and through the FCT Project CERN/FIS-PAR/0023/2019. 

\appendix

\section{Relating $(Y, \Upsilon)$ to $( \hat{Y}^I, \hat{\Upsilon} ) $ \label{relYU}}

In \cite{LopesCardoso:2000qm}, the authors considered a particular rescaling of the fields $(X^I, {\hat A})$. Here,
we will denote the resulting rescaled fields by $( \hat{Y}^I, \hat{\Upsilon} )$. The rescaling made use of the metric
factor $e^{g}$ in the 
four-dimensional space-time metric,
\bea
ds_4^2 = - e^{2g(R)} \, dt^2 + e^{- 2 g(R)} \, \left( dR^2 + R^2  \, \left( d \theta^2 + \sin^2 \theta \, d \varphi^2 \right) \right)\;,
\label{lin4db}
\eea
as well as of a $U(1)$ phase $h$, which here we identify with the phase of $w$,  $w = | w | \, h $, introduced in \eqref{eq:T-w}.
The rescaled fields $( \hat{Y}^I, \hat{\Upsilon} )$ are given by
\bea
\hat{Y}^I = e^{-g} \, {\bar h} \, X^I \;\;\;,\;\;\;  \hat{\Upsilon} = e^{-2g} \, {\bar h}^2 \, {\hat A} \;.
\eea
Comparing with \eqref{rescalXA}, we infer the relations
\bea 
Y^I = \lambda \, {\hat Y}^I  \;\;\;,\;\;\; \Upsilon =   \lambda^2  \, {\hat \Upsilon} \;,
\label{rescTU}
\eea
with
\bea
\lambda = \frac14 v_2 |w| e^g \;\;\;,\;\;\; v_2 =  e^{- 2 g} \, R^2 \;.
\eea
Using \eqref{Pgc}, we obtain
\bea
e^{- 2 g} =G_4 \, e^{- {\cal K} ({\hat Y})} \;,
\label{gky}
\eea
where $e^{- {\cal K} ({\hat Y})} $ takes the form \eqref{calKY}, with $Y^I$ replaced by ${\hat Y}^I$.
Combining this with \eqref{vupsky}, we get
\bea
e^{- {\cal K} ({\hat Y})} =  8 \,   \frac{e^{- {\cal K} (Y)} }{  R^2 \, \sqrt{- \Upsilon} } \;.
\eea
Inserting $Y^I = \lambda \, {\hat Y}^I $ into this equation, we infer
\bea
\lambda^2 = \frac{R^2 \sqrt{- \Upsilon } }{8} \;. 
\eea
Then,
\bea
\Upsilon =   \lambda^2 
 \, {\hat \Upsilon} =  \frac{R ^2 \sqrt{- \Upsilon } }{8} \, \hat{\Upsilon} \;,
 \label{upsups}
\eea
and hence
\bea
\sqrt{-\Upsilon } =-  \hat{\Upsilon} \, \frac{R^2 }{8} \;\;\;,\;\;\;
\Upsilon= - \hat{\Upsilon}^2 \, \frac{R^4 }{64 } \;.
\eea
Furthermore, 
\bea
\lambda = \frac{1}{\sqrt{8}} \, R \,  \left( - \Upsilon \right)^{1/4}  = \frac{1}{8} \, R^2 \, \sqrt{- \hat{\Upsilon} } 
\eea
and hence,
\bea 
Y^I = \frac{ R^2 \, \sqrt{- \hat{\Upsilon} } }{  8}  \, \hat{Y}^I  \;.
\label{YY}
\eea
Thus, equations \eqref{upsups} and \eqref{YY} relate $(Y, \Upsilon)$ to $( \hat{Y}^I, \hat{\Upsilon} ) $.

\section{First-order flow equations from Hamilton's principal function \label{sec:hpf}}

In Fefferman-Graham gauge \eqref{backgr}, and in the gauge $A_r^I = 0 , {\tilde A}_{ r  I} = 0$, 
the canonical momenta conjugate to $h_{tt}, v_2, Y^I, A_t^I, {\tilde A}_{t I}$, computed using the 
two-derivative bulk Lagrangian \eqref{lagHef} and the Gibbons-Hawking boundary term
\eqref{K}, take the form
\begin{align}
\pi^{tt} &
= \frac{\sqrt{-h} }{2 G_4} \, h^{tt} \partial_r v_2 \,, \nn \\
\pi_{v_2} 
& = \frac{\sqrt{-h}  }{2 G_4}  \left( \frac{\partial_r h_{tt}}{h_{tt}} 
+ \frac{\partial_r  v_2}{v_2} \right)\,,  \\
\pi^Y_I  &= 
-  \frac{\sqrt{-h} \,v_2 \,e^{\mathcal{K} (Y)}\, N_{IJ}}{G_4} \left( \partial_r \bar{Y}^J + e^{\mathcal{K}(Y)} \, \bar{Y}^J \, Y^L N_{KL}  \, \partial_r \bar{Y}^K \right) \,, \nn \\
\pi^t_I  &= 
-  \frac{\sqrt{-h} \,v_2 \, h_{tt} }{2} \left[
\left( N_{IJ} + R_{IK} N^{KL} R_{LJ} \right) \partial_r A_t^J - 2 R_{IJ} N^{JK} \partial_r {\tilde A}_{t K} \right] + 2 i \left( F_I - {\bar F}_I \right) \,, \nn \\
{\tilde \pi}^{t I }  &= 
-  \frac{\sqrt{-h} \,v_2 \, h_{tt} }{2} \left[
 4 N^{IJ}  \partial_r {\tilde A}_{t J} - 2 N^{IJ} R_{JK}  \partial_r A_t^K \right] - 2 i \left( Y^I - {\bar Y}^I \right) \,, \nn
\label{canomom}
\end{align} 
where $e^{- \mathcal{K}(Y)} = i \left( \bar{Y}^I \, F_I (Y) - {\bar F}_I (\bar{ Y}) \, Y^I \right) $. We note the constraint $\pi^Y_I Y^I =0$.
Inverting the above relations except the one for  $\pi^Y_I$, we obtain
\bea
\label{momv2gam}
\partial_r v_2 &=& \frac{2 G_4}{\sqrt{- h}} h_{tt} \pi^{tt} = \frac{2 G_4}{\sqrt{- h}} \pi^t_t \;, \nonumber\\
\partial_ r \gamma_{tt} &=&  \frac{2 G_4}{\sqrt{- h}} \gamma_{tt} \left( \pi_{v_2} - \frac{\pi^t_t}{v_2} \right) \;, \\
\partial_r A_t^I &=& 2 \frac{\sqrt{- h}}{v_2} N^{IJ} \left( \pi^t_J + \frac12 R_{JK} {\tilde \pi}^{t K} + N_{JK} \left( Y^K + {\bar Y}^K \right) \right) \;,
\nonumber\\
\partial_r {\tilde A}_{t I} &=&  \frac{\sqrt{- h}}{v_2}  \left( R_{IJ} N^{JK} \pi^t_K 
+ \frac12 \left( N_{IJ} + 
R_{IK} N^{KL} R_{LJ} \right)
{\tilde \pi}^{t J} + 2 \left( F_I + {\bar F}_I \right) \right)\;.
\nonumber
\eea
Next, we decompose the combination $Y^L N_{KL}  \, \partial_r \bar{Y}^K$ into real and imaginary parts,
\bea
Y^L N_{KL}  \, \partial_r \bar{Y}^K = {\cal R} + i {\cal I} \;,
\eea
where we note that 
\bea
e^{{\cal K}(Y)}  {\cal R} =\frac{e^{{\cal K}(Y)} }{2} \partial_r \left( Y^L N_{KL}  \, \bar{Y}^K \right) 
e^{- {\cal K}(Y)/2}
 \partial_r e^{{\cal K}(Y)/2} \;.
\eea
We then write the canonical momentum $\pi^Y_I $ as
\bea
\label{relpiY}
\pi^Y_I  &=&
 -  \frac{\sqrt{-h} \,v_2 \,e^{\mathcal{K}(Y)}\, N_{IJ}}{G_4} \left( \partial_r \bar{Y}^J + e^{\mathcal{K}(Y)} \, \bar{Y}^J \left(
 {\cal R} + i {\cal I} \right) \right)\nonumber\\
&=&  -  \frac{\sqrt{-h} \,v_2 \, e^{\mathcal{K}(Y)/2} \, N_{IJ}}{G_4} \left( \partial_r \left( e^{\mathcal{K}(Y)/2} \bar{Y}^J \right) 
+ i 
e^{3 \mathcal{K}(Y)/2} \, \bar{Y}^J  {\cal I}  \right) \;.
\eea
In general, this relation cannot be inverted due to the constraint $\pi^Y_I Y^I =0$. However, 
we observe that when 
\bea
 {\cal I} = 0 \;,
 \label{condI}
 \eea
 we may invert the relation \eqref{relpiY}. The condition \eqref{condI} is, for instance, satisfied by BPS solutions, see \eqref{zbzi}.
 Thus, in the following, we will restrict to solutions that satisfy \eqref{condI}.
 
 Inserting \eqref{condI} into the above equation for $\pi^Y_I$ gives
 \bea
\pi^Y_I  &=&  -  \frac{\sqrt{-h} \,v_2 \, e^{\mathcal{K}(Y)/2} \, N_{IJ}}{G_4} \partial_r \left( e^{\mathcal{K}(Y)/2} \bar{Y}^J \right) \;,
\eea
and hence
\bea
\partial_r \left( e^{\mathcal{K}(Y)/2} \bar{Y}^I \right) = - \frac{G_4}{ \sqrt{-h} \,v_2 \, e^{\mathcal{K}(Y)/2} } N^{IJ} \pi^Y_J \;,
\label{momYrel}
\eea
subject to \eqref{condI}.

Next, following \cite{Castro:2018ffi}, we write 
Hamilton's principal function ${\cal S}$, which only depends
on coordinates and not on canonical momenta, as follows,
\bea
{\cal S} ( h_{tt}, v_2, Y^I, {\bar Y}^I, A_t^I, {\tilde A}_{t I}) = U (h_{tt}, v_2, Y^I, {\bar Y}^I) + 
\int dt \left( - q_I A_t^I + p^I {\tilde A}_{t I} \right) \;.
\eea
This gives
\bea
\pi^t_I = \frac{\delta {\cal S} }{\partial A_t^I }= - q_I  \;\;\;,\;\;\; {\tilde \pi}^{t I}  = \frac{\delta {\cal S} }{\partial {\tilde A}_{t I}  } = p^I \;.
\eea
We restrict ourselves to the case when $U$ does not depend on $h_{tt}$,
\bea
U (v_2, Y^I, {\bar Y}^I) = \frac{1}{G_4} \int dt \sqrt{- h} \, W (v_2, Y^I, {\bar Y}^I) \;,
\eea
where $W$ is real.
Using 
\bea
\pi^{tt} =  \frac{\delta {\cal S} }{\delta h_{tt}} = - \frac{1}{2 G_4 \sqrt{- h}} \, W \;\;\;,\;\;\;
\pi_{v_2} = \frac{\delta {\cal S} }{\delta v_2} = \frac{ \sqrt{- h} }{G_4} \, \frac{ \partial W }{\partial v_2} \;,
\eea
and inserting this into the relations \eqref{momv2gam} gives
\bea
\partial_r v_2 &=&W \;, \nonumber\\
\partial_ r h_{tt} &=&  2 h_{tt} \left( \frac{ \partial W}{\partial v_2} - \frac12 \frac{W}{v_2} \right) \;.
\label{flowvg}
\eea
Finally, using
\bea
\pi^{Y}_I =  \frac{\delta {\cal S} }{\delta Y^I} =  \frac{ \sqrt{- h} }{G_4} \, \frac{ \partial W }{\partial Y^I} \;,
\eea
we infer
\bea
\partial_r \left( e^{\mathcal{K}/2} \bar{Y}^I \right) = - \frac{1}{ v_2 \, e^{\mathcal{K}(Y)/2} } N^{IJ} \, \frac{ \partial W }{\partial Y^J} \;.
\label{flowY}
\eea
Inserting this relation into \eqref{condI} yields
\bea
 {\rm Im} \left( Y^I \, \frac{ \partial W }{\partial Y^I} \right) = 0 \;.
 \label{Icond}
\eea
We take 
\bea
W (v_2, Y, {\bar  Y}) = 2 \left( \sqrt{v_2} - e^{ {\cal K} (Y)/2} \, {\rm Re}  Z( Y)  \right)  \;,
\label{Wsupv}
\eea
where
\bea
Z(Y) =  p^I F_I(Y) - q_I Y^I \;.
\eea
Then
\bea
 {\rm Im} \left( Y^I \, \frac{ \partial W }{\partial Y^I} \right) = 0 \Longleftrightarrow  Z(Y) = {\bar Z} ({\bar Y}) \;.
 \label{zbzi}
 \eea
 
 Thus, summarizing, we obtained the following set of flow equations in Fefferman-Graham coordinates,
 \bea
\partial_r v_2 
&=&W \;, \nonumber\\
\partial_ r h_{tt} 
&=&  2 h_{tt} \left( \frac{ \partial W}{\partial v_2} - \frac12 \frac{W}{v_2} \right) \;, \nonumber\\
\partial_r  \left( e^{\mathcal{K} (Y)/2} \, \bar{Y}^I \right) 
&=& - \frac{1}{ v_2 \, e^{\mathcal{K}(Y)/2} } N^{IJ} \, \frac{ \partial W }{\partial Y^J} \;,
\label{flwf}
\eea
with
\bea
W (v_2, Y, {\bar  Y}) = 2 \left( \sqrt{v_2} - e^{ {\cal K} (Y)/2} \, {\rm Re}  Z( Y)  \right)  \;,
\label{WZ}
\eea
where
\bea
Z(Y) = p^I F_I (Y) - q_I \, Y^I = p^I F_I ({\bar Y}) - q_I \, {\bar Y}^I =  {\bar Z} ({\bar Y}) \;.
\label{condZbZ}
\eea
These flow equations are solved by BPS solutions, as we will verify in Appendix \ref{sec:flbp}.
In Appendix \ref{sec:nBPS} we will show that a 
particular deformation of these flow equations 
encodes the nAdS$_2$ attractor mechanism \cite{Larsen:2018iou} when working infinitesimally away from the horizon.

\section{BPS flow equations in Fefferman-Graham coordinates \label{sec:flbp}}

In the following, we set $G_4 = 1$ for convenience, 
and we show that the flow equations for single centre BPS solutions in four dimensions at the two-derivative level take the form 
\eqref{flwf}, \eqref{WZ}, \eqref{condZbZ} when written in Fefferman-Graham coordinates.

The standard form of 
the BPS flow equations \cite{Ferrara:1995ih,Ferrara:1997tw}
uses 
 the four-dimensional line element \eqref{lin4db}. These flow equations, when formulated in terms of the scalar fields ${\hat Y}^I$ defined in Appendix \ref{relYU},  take the form (see, for instance, eq. (671) in \cite{Cardoso:2019wwf}),
\bea
R^2 \partial_R g&=& e^{2g}  \, {\rm Re} Z({\hat Y} )  \;, \nonumber\\
R^2 \, \partial_R {\hat Y}^I &=&  N^{IJ} \frac{\partial}{\partial \bar{\hat Y}^J} {\bar Z} ({\bar {\hat Y}}) \;, \nonumber\\
{\rm Im} Z ({\hat Y}) &=& 0 \;, \nonumber\\
{\hat A}_R &=& 0 \;,
\label{floww} 
\eea
where
\bea
Z({\hat Y}) &=&  p^I F_I({\hat Y}) - q_I {\hat Y}^I \;, \\
{\hat A}_R &=& - \ft12 e^{2g} \left[ \left(F_I ({\hat Y})  - {\bar F}_I (\bar{\hat Y} ) \right) 
 \partial_R \left({\hat Y}^I - \bar{\hat Y}^I \right) -  \left({\hat Y}^I - \bar{\hat Y}^I \right) \partial_R  \left(F_I ({\hat Y})  - {\bar F}_I (\bar{\hat Y} ) \right)  \right]  \;. \nonumber
\eea
Note that when written in terms of the scalar fields ${Y}^I$,
BPS solutions satisfy the condition
\bea
e^{- {\cal K}(Y)} =  {\rm Re} Z(Y) > 0\;.
\label{phi30}
\eea
This can be seen as follows. Using \eqref{YY}, we obtain
\bea
e^{- {\cal K}(Y)} = e^{- {\cal K}({\hat Y})} \,\frac{R^4}{64}  \,  \left( - \hat{\Upsilon} \right)   \;.
\label{kkrel}
\eea
Using that for a BPS solution \cite{LopesCardoso:2000qm}
\bea
 \sqrt{- \hat{\Upsilon}}  = 8 | \partial_R g | \;,
 \eea
 we obtain from \eqref{kkrel},
 \bea
e^{- {\cal K}(Y)} = e^{- {\cal K}({\hat Y})} \,\frac{R^4}{8} \,  \sqrt{- \hat{\Upsilon}}  \;  | \partial_R g | = e^{- 2g} \,\frac{R^4}{8} \,  \sqrt{- \hat{\Upsilon}} \,  
| \partial_R g | \;,
\eea
where we used \eqref{gky}. Using the first flow equation in \eqref{floww} we infer
\bea
e^{- {\cal K}(Y)} =  \frac{R^2}{8} \,  \sqrt{- \hat{\Upsilon}} \; \vert {\rm Re} Z({\hat Y})  \vert = \vert {\rm Re} Z(Y)  \vert \;,
\eea
where we used \eqref{YY} once more. Finally, demanding 
$\partial_R g > 0$ along the BPS flow \cite{Ferrara:1997tw}, and hence 
${\rm Re} Z({\hat Y}) >0$, 
we obtain \eqref{phi30}.

Note that, in general, \eqref{phi30} does not imply that the $Y^I$ are constant along a BPS flow. 
Double-extreme BPS black holes are solutions where 
the $Y^I$ are constant along the BPS flow. These are
discussed in section \ref{sec:de}.

Let us now verify that the standard flow equations \eqref{floww} become the flow equations \eqref{flwf} in Fefferman-Graham coordinates.
To this end, we will use the following relation, which is a consequence of \eqref{rescTU}.
\bea
e^{ {\cal K}(Y)/2} \, Z(Y) = e^{ {\cal K}({\hat Y}) /2} \, Z({\hat Y}) \;.
\label{kzkz}
\eea
Using $\partial_r = e^g \, \partial_R$ as well as 
$v_2 = e^{-2g} R^2$, we obtain
\bea
\partial_r v_2 = e^g \partial_R \left( e^{-2g} R^2 \right) = - 2 e^{-g} \partial_R g \, R^2 + 2 e^{-g} R =  - 2 e^{-g} \partial_R g \,  R^2 + 2 \sqrt{v_2} \;.
\eea
Using the first equation of \eqref{floww}, this becomes
\bea
\partial_r v_2 = - 2 e^{g} \, {\rm Re} Z({\hat Y}) + 2 \sqrt{v_2} \;.
\eea
Using first \eqref{gky} and then \eqref{kzkz}, we obtain
\bea
\partial_r v_2 = - 2  e^{ {\cal K}(Y)/2} \, {\rm Re} Z(Y) 
 + 2 \sqrt{v_2} = W\;.
\eea
Next, let us verify the flow equation for $h_{tt}$,
\bea
\partial_ r h_{tt} &=&  e^g \partial_R \left( - e^{2g} \right) =  - 2 e^{5g}\,  \frac{{\rm Re} Z({\hat Y})}{R^2} \nonumber\\
&=& \frac{2}{v_2} h_{tt} \, e^g \,  {\rm Re} Z({\hat Y}) = \frac{2}{v_2} h_{tt} \, e^{{\cal K}({\hat Y}) /2}  \, {\rm Re} Z({\hat Y}) \nonumber\\
&=& \frac{2}{v_2} h_{tt} \, e^{{\cal K}(Y)/2}  \, {\rm Re} Z(Y) = 2 h_{tt} \left( \frac{ \partial W}{\partial v_2} - \frac12 \frac{W}{v_2} \right) \;.
\eea
Finally, we verify the flow equation for $Y^I$, 
\bea
\partial_r \left( e^{\mathcal{K}(Y)/2} \,  \bar{Y}^I \right) &=& e^g \partial_R \left( e^{\mathcal{K}({\hat Y}) /2} \, \bar{{\hat Y}}^I \right) = 
e^g  \partial_R \left( e^{\mathcal{K} ({\hat Y}) /2} \right) \bar{\hat Y}^I  + e^{\mathcal{K} ({\hat Y}) /2}
\frac{e^g}{R^2} \, N^{IJ} \frac{\partial}{\partial {{\hat Y}}^J} {Z} ({{\hat Y}}) \nonumber\\
&=& e^g  \partial_R \left( e^g \right) \bar{{\hat Y}}^I  + \frac{1}{v_2} \, N^{IJ} \frac{\partial}{\partial {{\hat Y}}^J} { Z} ({\hat Y}) \nonumber\\
&=& \frac{e^{2g}}{R^2}  \, \partial_R g \,  R^2 \,  \bar{\hat Y}^I + \frac{1}{v_2} \, N^{IJ} \frac{\partial}{\partial {{\hat Y}}^J} { Z} ({\hat Y}) \nonumber\\
&=& \frac{1}{v_2} \, \bar{Y}^I + \frac{1}{v_2} \, N^{IJ} \frac{\partial}{\partial {Y}^J} {Z} ({Y}) \;.
\label{flY}
\eea
This we compare with 
\bea
 - \frac{1}{ v_2 \, e^{\mathcal{K}(Y)/2} } N^{IJ} \, \frac{ \partial W }{\partial Y^J} = \frac{2}{v_2 \, e^{\mathcal{K}(Y)/2} }  N^{IJ} \,
 \left( \frac{\partial e^{\mathcal{K}(Y)/2}}{\partial Y^J}\right) {\rm Re} Z(Y)  + \frac{1}{v_2} \, N^{IJ} \frac{\partial}{\partial {Y}^J} {Z} ({Y}) \;.
 \nonumber\\
\eea
Next, using $e^{- {\cal K} (Y) } = - {\bar Y}^I N_{IJ} Y^J$, we write
\bea
 \frac{\partial e^{\mathcal{K}(Y)/2}}{\partial Y^J} = - \frac{ e^{3 \mathcal{K}(Y)/2} }{2} \,
 \frac{\partial e^{\mathcal{- K}(Y)}}{\partial Y^J} =  \frac{ e^{3 \mathcal{K}(Y)/2} }{2} \,N_{JK} {\bar Y}^K \;,
  \eea
so that 
\bea
 - \frac{1}{ v_2 \, e^{\mathcal{K}(Y)/2} } N^{IJ} \, \frac{ \partial W }{\partial Y^J} = \frac{e^{\mathcal{K}(Y)}}{v_2 }  \, {\bar Y}^I \,
   {\rm Re} Z(Y)  + \frac{1}{v_2} \, N^{IJ} \frac{\partial}{\partial {Y}^J} {Z} ({Y}) \;.
\eea
Then, using \eqref{phi30}, 
we obtain
\bea
 - \frac{1}{ v_2 \, e^{\mathcal{K}(Y)/2} } N^{IJ} \, \frac{ \partial W }{\partial Y^J} = \frac{1}{v_2 }  \, {\bar Y}^I \,
    + \frac{1}{v_2} \, N^{IJ} \frac{\partial}{\partial {Y}^J} {Z} ({Y}) \;.
\eea
Thus, from \eqref{flY} we infer,
\bea
\partial_r \left( e^{\mathcal{K}(Y)/2} \bar{Y}^I \right) = - \frac{1}{ v_2 \, e^{\mathcal{K}(Y)/2} } N^{IJ} \, \frac{ \partial W }{\partial Y^J} \;.
\eea

Finally, we note that the condition ${\rm Im} Z ({\hat Y}) =0$ in \eqref{floww} is equivalent to 
\eqref{condZbZ} by virtue of \eqref{rescTU}, and that the condition ${\hat A}_R =0$, when written in terms of the fields $Y^I$, is satisfied by virtue of
\eqref{condI}.

\section{nAdS$_2$ from deformed BPS first-order flow equations \label{sec:nBPS}}

At the two-derivative level, non-extremal single centre static black hole solutions in four dimensions 
are obtained from extremal single centre static black hole solutions by turning on a deformation parameter $\mu$.
The nAdS$_2$ attractor mechanism \cite{Larsen:2018iou} posits that to first order in $\mu$, the near-horizon behaviour of the 
metric and of the scalar fields of the non-extremal black hole solution is governed by the near-horizon behaviour of the extremal
black hole solution. 
We will refer to the deformed solution as a 
nAdS$_2$ deformation of an extremal black hole solution. In the following, we will show that at first order in $\mu$ and infinitesimally
away from the horizon, the nAdS$_2$ deformation
of BPS black hole solutions is governed by first-order
flow equations that are a deformation of 
the flow equations \eqref{flwf}.

The most general line element for a static, spherically symmetric non-extremal black hole solution at the two-derivative level in four dimensions takes the form
\cite{Larsen:2018iou},
\bea
ds_4^2 = - e^{2 p(R)} \, f(R) \, dt^2 + e^{-2p(R)} \,  f^{-1}(R) \, d R^2 + e^{-2p(R)}  \, d\Omega_2^2 \;,\; f(R) = R^2 - \mu^2 > 0,
\eea
where $R > \mu > 0$.
In these coordinates, the outer horizon of the non-extremal black hole is at $R = \mu$. A BPS black hole in ungauged $N=2$ supergravity theories
is an extremal black hole that 
satisfies $\mu = 0$ as well as
\bea
e^{-2p(R)} = e^{- 2 g(R)} \, R^2 = i \left( \bar{\hat{Y}}^I \, F_I (\hat{Y}) - {\bar F}_I (\bar{\hat Y}) \, {\hat Y}^I \right) \, R^2 \;,
\eea
where the ${\hat Y}^I$ are determined in terms of harmonic functions $(H^I, H_I)$ according to \eqref{hatYHH}.

Upon changing the radial coordinate as $R \rightarrow  R +  \mu $, the above line element takes the form
\bea
ds_4^2 = - e^{2 p(R)} \, f(R) \, dt^2 + e^{-2p(R)} \,  f^{-1}(R) \, d R^2 + e^{-2p(R)}  \, d\Omega_2^2 \;\;\;,\;\;\;  f(R) = R^2  + 2 \mu R > 0, 
\nonumber\\
\label{nonextlin}
\eea
where now $R > 0$, and the outer horizon of the non-extremal black hole is at $R = 0$.
This line element can be brought into the form \eqref{backgr} by performing the change of coordinates
\bea
 \frac{\partial}{\partial r} = e^{p(R)} \, \sqrt{f(R)} 
\, \frac{\partial}{\partial R} \;,
\label{rRr}
\eea
in which case
\bea
v_2 (r) = e^{- 2 p(R)}  \vert_{R = R(r)} \;\;\;,\;\;\; h_{tt} (r) = - e^{2p(R)} \, f(R) \vert_{R = R(r)} 
\;.
\label{v2h}
\eea
The solution \eqref{v2h} is supported by complex scalar fields $Y^I (r)$, c.f. \eqref{rescalXA}.

 In the following, we will derive the nAdS$_2$ attractor mechanism for non-extremal black holes \cite{Larsen:2018iou} 
 by demanding that at lowest order in $\mu$ and $R$,
$v_2, h_{tt} $ and $Y^I$ satisfy flow equations that are a particular deformation of the flow equations given in 
\eqref{flwf}, as follows. Using \eqref{rRr}, the radial derivative $\partial_r \phi$ of a field $\phi$
can be expressed in terms of $\partial_R \phi$ as $\partial_r \phi = e^{p(R)} \, \sqrt{f(R)} \, \partial_R \phi$. Now multiply this expression
by $\frac12 (\partial_R f )/\sqrt{f}$, which equals $1$ when $\mu = 0$,
thereby obtaining the combination $\frac12 e^{p(R)} \, (\partial_R f ) \, 
\partial_R \phi$. Proceeding in this manner with the flow equations  \eqref{flwf}, we are led to consider
the following deformed flow equations,
\bea
\frac12 e^p \, (\partial_R f )\, 
\partial_R v_2 
&=&W \;, \nonumber\\
\frac12  e^p \, (\partial_R f )
 \, \partial_ R h_{tt} 
&=&  2 h_{tt} \left( \left( 1 + \frac{\mu^2}{f} \right) \frac{ \partial W}{\partial v_2} - \frac12 \frac{W}{v_2} \right) \;, \nonumber\\
\frac12  e^p \, 
(\partial_R f ) \, \partial_R  \left( e^{\mathcal{K} (Y)/2} \, \bar{Y}^I \right) 
&=& - \frac{1}{ v_2 \, e^{\mathcal{K}(Y)/2} } N^{IJ} \, \frac{ \partial W }{\partial Y^J} \;,
\label{flwfmod}
\eea
where $W$ is given by \eqref{WZ}, and where we impose \eqref{condZbZ}. 
Note that we have added a term proportional to $\mu^2/f$ on the right hand side of the flow equation for $h_{tt}$,
and that $\mu^2/f$ is of order zero in the small quantities $\mu$ and $R$.
We now solve the flow equations \eqref{flwfmod} to lowest order in $\mu$ and in $R$.

Working infinitesimally away from the horizon $R=0$ and to first order in 
the perturbation parameter $\mu$, we expand the metric factor $e^{-2p(R)}$ in \eqref{nonextlin} and the complex scalar fields $Y^I$ as 
\bea
\label{expconst}
e^{-2p(R)} &=& A^2 + \alpha \, \mu + 2 B R  
+  {\cal O} (R^2, \mu^2, \mu R)  \;, 
\nonumber\\
Y^I (R) &=& Y^I_0 +  {y}^I  \mu + Z^I R
+  {\cal O} (R^2, \mu^2, \mu R)  \;, 
\eea
where $A^2, \alpha, B, Y^I_0, y^I, Z^I$ denotes constants.
In this expansion, terms proportional to $\mu R$, $\mu^2$ or $R^2$ are taken to be of the same order.
When setting $\mu =0$, the resulting solution describes the near-horizon solution of a BPS black hole and hence,
the constants $A^2$ and $Y_0^I$ are determined by the BPS attractor equations \eqref{2dBPS}. The remaining constants in 
\eqref{expconst} are determined through the deformed flow equations \eqref{flwfmod}, as follows. First, we list the following expansions, for later use (we
take $A > 0$),
\bea
e^{-p(R)} &=& A \left( 1 + \frac{\alpha \, \mu }{2 A^2}  +  \frac{B }{A^2} R   \right) +  {\cal O} (R^2, \mu^2, \mu R)  \;, \nonumber\\
e^{p(R)} &=& \frac{1}{A}  \left( 1 - \frac{\alpha \, \mu }{2A^2}  -  \frac{B }{A^2} R    \right) +  {\cal O} (R^2, \mu^2, \mu R)\;.
\label{expf}
\eea
First we analyze the flow equation for $v_2$, 
\bea
\frac12  e^{p(R)} (\partial_R f ) \partial_R v_2 = 
e^{p(R)} (R + \mu) \, \partial_R e^{-2p(R)} = W \;.
\eea
At first order in $\mu$ and $R$, the left hand side gives
\bea
\frac{2B}{A}  
 (R+ \mu) \;,
 \label{prv2}
\eea
 while the right hand side gives
 \bea
 \label{Wexp}
 W &=& 2 \left(  A + \frac{ \alpha \mu }{2 A}  +  \frac{B }{A} R  - e^{ - {\cal K} (Y_0)/2} -  \partial_I \left( e^{ {\cal K} (Y)/2} \, {\rm Re} Z(Y) \right) \vert_{Y_0} \,  
 \left( Z^I R +  {y}^I  \mu \right) \right. \nonumber\\
&& \left. \qquad   - \partial_{\bar I} \left( e^{ {\cal K} (Y)/2} \, {\rm Re} Z(Y) \right) \vert_{Y_0} \,  \left( {\bar Z} R +  {{\bar y}}^I  \mu \right)
  \right)  \;.
   \eea
 Using $e^{ - {\cal K} (Y)} = - N_{IJ} {\bar Y}^I Y^J $, we obtain
 \bea
\label{derWexp}
&& \partial_I \left( e^{ {\cal K} (Y)/2} \, {\rm Re} Z(Y) \right) \vert_{Y_0} = \left[ \frac12 N_{IJ} {\bar Y}^J e^{ 3 {\cal K} (Y)/2} \, {\rm Re} Z(Y)
 + \frac 12 e^{ {\cal K} (Y)/2} \, \left( p^J F_{JI} - q_I \right) \right] \vert_{Y_0} \nonumber\\
 &=&  \frac 12 e^{ {\cal K} (Y_0)/2} \left[ N_{IJ} {\bar Y}^J  + p^J F_{JI} - q_I \right]  \vert_{Y_0}  = 0 \;,
 \eea
 where we used the BPS relation $e^{-{\cal K}(Y_0)} = {\rm Re} Z(Y_0) $ and the BPS attractor equations \eqref{2dBPS} for $Y_0^I$.
 This results in
 \bea
 W =  \frac{\alpha \mu }{ A}  +  2 \frac{B }{A} R  \;,
\eea
where we used the BPS relation $A^2 = e^{ - {\cal K} (Y_0)}$. Comparing this with \eqref{prv2} determines $\alpha$ to equal
\bea
\alpha = 2 B \;.
\label{vaB}
\eea

Next, we consider the flow equation for $h_{tt}$,
\bea
\frac12  e^{p(R)} (\partial_R f )
\partial_R h_{tt} = 2 h_{tt} \,\frac{(R + \mu)}{f}  \left( \left( \partial_R  e^p \right) f  + \frac{e^p}{2} \partial_R f \right) 
= 
 2 h_{tt} \left( \left( 1 + \frac{\mu^2}{f} \right) \frac{ \partial W}{\partial v_2} - \frac12 \frac{W}{v_2} \right) \;,
 \nonumber\\
\eea
which yields
\bea
\frac{(R + \mu)}{f}  \left(  \left( \partial_R e^p \right) f  + \frac{e^p}{2}  \partial_R f \right)= \left( 1 + \frac{\mu^2}{f} \right) \frac{ \partial W}{\partial v_2} - \frac12 \frac{W}{v_2} \;.
\label{flowhttsimp}
 \eea
Using \eqref{vaB}, the left hand side gives
\bea
&&- \frac{B}{A^3} \left( R + \mu \right) + \frac{1}{A}  \left( 1 - \frac{\alpha \mu }{2A^2}  -  \frac{B }{A^2} R    \right) \left( 1 + \frac{\mu^2}{f} \right) \;,
\label{lhshf}
\eea
which precisely equals the combination on the right hand side of \eqref{flowhttsimp}.

Finally, we turn to the flow equation for ${\bar Y}^I$,
\bea
\frac12  e^p (\partial_R f ) \partial_R \left( e^{\mathcal{K} (Y)/2} \, \bar{Y}^I \right) =
e^p (R + \mu) \partial_R \left( e^{\mathcal{K}(Y)/2} {\bar Y}^I \right)
= - \frac{1}{ v_2 \, e^{\mathcal{K}(Y)/2} } N^{IJ} \, \frac{ \partial W }{\partial Y^J} .
\label{Yfl}
\eea
To first order in $\mu, R$ we obtain
\bea
e^{\mathcal{K}(Y)/2} {\bar Y}^I &=& e^{\mathcal{K}(Y_0)/2} \left(  {\bar Y}^I_0 + {\bar Z}^I R + {\bar y}^I \mu \right) \\
&& + \frac12 e^{3 \mathcal{K}(Y_0)/2} {\bar Y}^I_0 \,
 N_{PQ}  \vert_{Y_0}  \left( {\bar Y}_0^P (  Z^Q R + y^Q \mu ) + Y_0^P
( {\bar  Z}^Q R + {\bar y}^Q \mu ) \right) \;, \nonumber
\eea
and hence,
\bea
\partial_R \left( e^{\mathcal{K}(Y)/2} {\bar Y}^I \right) = e^{\mathcal{K}(Y_0)/2} {\bar Z}^I
+ \frac12 e^{3 \mathcal{K}(Y_0)/2} {\bar Y}^I_0 \,
 N_{PQ}  \vert_{Y_0}  \left( {\bar Y}_0^P Z^Q  + Y_0^P {\bar  Z}^Q  \right) \;.
\eea
We then obtain for the left hand side of \eqref{Yfl},
\bea
\frac12  e^p (\partial_R f ) \partial_R \left( e^{\mathcal{K}(Y)/2} {\bar Y}^I \right) &=& \frac{1}{A} ( R + \mu) \left[ e^{\mathcal{K}(Y_0)/2} {\bar Z}^I
\right. \nonumber\\
&& \left. + \frac12 e^{3 \mathcal{K}(Y_0)/2} {\bar Y}^I_0 \,
 N_{PQ}  \vert_{Y_0}  \left( {\bar Y}_0^P Z^Q  + Y_0^P {\bar  Z}^Q  \right) \right] \;.
\eea
Next, we compute the right hand side of \eqref{Yfl}. Expanding around $Y_0^I$ and 
using \eqref{derWexp}, we obtain,
\bea
\partial_I W = - 2 \partial_I \left( e^{\mathcal{K}(Y)/2}  \, {\rm Re} Z \right)  
=
- ({\bar Y} - {\bar Y}_0)^J \, N_{IJ}\vert_{Y_0}  \, e^{\mathcal{K}(Y_0)/2} \;.
\eea
Then, the flow equation
\eqref{Yfl} becomes, to first order in $\mu, R$,
\bea
 \frac{1}{A} ( R + \mu) \left[ e^{\mathcal{K}(Y_0)/2} {\bar Z}^I
 + \frac12 e^{3 \mathcal{K}(Y_0)/2} {\bar Y}^I_0 \,
 N_{PQ}  \vert_{Y_0}  \left( {\bar Y}_0^P Z^Q  + Y_0^P {\bar  Z}^Q  \right) \right] &=&  \frac{1}{A^2} \left( {\bar Z}^I R + {\bar y}^I \mu \right) 
\;. \nonumber\\
\eea
This equation is satisfied if we demand
\bea
y^I = Z^I \;\;\;,\;\;\;  Y_0^P \, N_{PQ}  \vert_{Y_0} {\bar  Z}^Q = 0 \;,
\eea
in which case
\bea
Y^I = Y^I_0 + Z^I (R + \mu) \;\;\;,\;\;\;  Y_0^P \, N_{PQ}  \vert_{Y_0} \left( Y^Q - Y_0^Q \right) = 0 \;.
\eea
Note that the latter condition is consistent with the condition \eqref{YdY} derived from the analysis of first-order perturbations.

Summarizing, for the expansions \eqref{expconst} we obtain at first order in $\mu, R$,
\bea
e^{-2p(R)} = A^2  + 2 B (R + \mu) 
\; \; ,\;\; Y^I (R) = Y^I_0 +  Z^I  (R + \mu) \;\;,\;\; Y_0^P \, N_{PQ}  \vert_{Y_0} {\bar  Z}^Q = 0 \;.
\eea
At this order, the expressions for $e^{-2p(R)}$ and $Y^I (R)$ can also be written as
\bea
e^{-2p(R)} &=& A^2  +  (R + \mu)  \, \frac{\partial e^{-2p_{\rm BPS}}(R)}{\partial R} \vert_{R=0} \;,
\nonumber\\
 Y^I (R) &=& Y^I_0 +   (R + \mu)  \frac{\partial Y_{\rm BPS}^I(R)}{\partial R} \vert_{R=0} \;,
\label{pY0}
\eea
where the metric function $p_{\rm BPS}(R)$ and the scalar fields  $Y_{\rm BPS}^I(R)$ are those that describe interpolating BPS solutions.
At the horizon $R=0$, we obtain the horizon values
\bea
e^{-2p(0)} &=& A^2  +  \mu  \, \frac{\partial e^{-2p_{\rm BPS}}(R)}{\partial R} \vert_{R=0} \;,
\nonumber\\
 Y^I (0) &=& Y^I_0 +  \mu \,   \frac{\partial Y_{\rm BPS}^I(R)}{\partial R} \vert_{R=0} \;,
\eea
and the physical scalar fields $z^A= Y^A/Y^0$ behave as
\bea
z^A (0) = z^A_0 + \mu \, \frac{\partial z_{\rm BPS} ^A (R)}{\partial R} \vert_{R=0} \;,
\eea
where  $z^A_0$ denotes the BPS value at the horizon. 
Thus, to first order in $\mu$, the horizon values $e^{-2p(0)}, Y^I(0)$ and $z^A(0)$ equal the BPS values 
$e^{-2p_{\rm BPS}(\mu)}, Y_{\rm BPS}^I(\mu)$ and $z_{\rm BPS}^A(\mu)$. This is precisely the
nAdS$_2$ attractor behaviour of the metric factor and of the scalar fields \cite{Larsen:2018iou}.
Note that since the BPS solution $Y_{\rm BPS}^I(R)$ satisfies \eqref{condZbZ}, also
 \eqref{pY0} satisfies the condition \eqref{condZbZ}.

 \providecommand{\href}[2]{#2}\begingroup\raggedright\endgroup


\begin{thebibliography}{10}

\bibitem{Gibbons:1982fy}
G.~W. Gibbons and C.~M. Hull, {\it {A Bogomolny Bound for General Relativity
  and Solitons in N=2 Supergravity}},  {\em Phys. Lett. B} {\bf 109} (1982)
  190--194.

\bibitem{Gibbons:1985bz}
G.~W. Gibbons, {\it {Solitons and black holes in four-dimensions,
  five-dimensions}},  \href{http://xxx.lanl.gov/abs/1903.04942}{{\tt
  1903.04942}}.

\bibitem{Spradlin:1999bn}
M.~Spradlin and A.~Strominger, {\it {Vacuum states for AdS(2) black holes}},
  {\em JHEP} {\bf 11} (1999) 021,
  [\href{http://xxx.lanl.gov/abs/hep-th/9904143}{{\tt hep-th/9904143}}].

\bibitem{Ferrara:1995ih}
S.~Ferrara, R.~Kallosh, and A.~Strominger, {\it {N=2 extremal black holes}},
  {\em Phys. Rev. D} {\bf 52} (1995) R5412--R5416,
  [\href{http://xxx.lanl.gov/abs/hep-th/9508072}{{\tt hep-th/9508072}}].

\bibitem{Ferrara:1996dd}
S.~Ferrara and R.~Kallosh, {\it {Supersymmetry and attractors}},  {\em Phys.
  Rev. D} {\bf 54} (1996) 1514--1524,
  [\href{http://xxx.lanl.gov/abs/hep-th/9602136}{{\tt hep-th/9602136}}].

\bibitem{Ferrara:1997tw}
S.~Ferrara, G.~W. Gibbons, and R.~Kallosh, {\it {Black holes and critical
  points in moduli space}},  {\em Nucl. Phys. B} {\bf 500} (1997) 75--93,
  [\href{http://xxx.lanl.gov/abs/hep-th/9702103}{{\tt hep-th/9702103}}].

\bibitem{Goldstein:2005hq}
K.~Goldstein, N.~Iizuka, R.~P. Jena, and S.~P. Trivedi, {\it
  {Non-supersymmetric attractors}},  {\em Phys. Rev. D} {\bf 72} (2005) 124021,
  [\href{http://xxx.lanl.gov/abs/hep-th/0507096}{{\tt hep-th/0507096}}].

\bibitem{Goldstein:2014gta}
K.~Goldstein, V.~Jejjala, and S.~Nampuri, {\it {Hot Attractors}},  {\em JHEP}
  {\bf 01} (2015) 075, [\href{http://xxx.lanl.gov/abs/1410.3478}{{\tt
  1410.3478}}].

\bibitem{Castro:2018ffi}
A.~Castro, F.~Larsen, and I.~Papadimitriou, {\it {5D rotating black holes and
  the nAdS$_{2}$/nCFT$_{1}$ correspondence}},  {\em JHEP} {\bf 10} (2018) 042,
  [\href{http://xxx.lanl.gov/abs/1807.06988}{{\tt 1807.06988}}].

\bibitem{Castro:2019vog}
A.~Castro and B.~M\"uhlmann, {\it {Gravitational anomalies in
  nAdS$_2$/nCFT$_1$}},  {\em Class. Quant. Grav.} {\bf 37} (2020), no.~14
  145017, [\href{http://xxx.lanl.gov/abs/1911.11434}{{\tt 1911.11434}}].

\bibitem{deAlfaro:1976vlx}
V.~de~Alfaro, S.~Fubini, and G.~Furlan, {\it {Conformal Invariance in Quantum
  Mechanics}},  {\em Nuovo Cim. A} {\bf 34} (1976) 569.

\bibitem{Larsen:2018iou}
F.~Larsen, {\it {A nAttractor mechanism for nAdS$_{2}$/nCFT$_{1}$ holography}},
   {\em JHEP} {\bf 04} (2019) 055,
  [\href{http://xxx.lanl.gov/abs/1806.06330}{{\tt 1806.06330}}].

\bibitem{Castro:2021wzn}
A.~Castro and E.~Verheijden, {\it {Near-AdS$_2$ Spectroscopy: classifying the
  spectrum of operators and interactions in $\mathcal{N} = 2$ 4D
  supergravity}},  \href{http://xxx.lanl.gov/abs/2110.04208}{{\tt 2110.04208}}.

\bibitem{LopesCardoso:2000qm}
G.~L. Cardoso, B.~de~Wit, J.~Kappeli, and T.~Mohaupt, {\it {Stationary BPS
  solutions in N=2 supergravity with $R^2$ interactions}},  {\em JHEP} {\bf 12}
  (2000) 019, [\href{http://xxx.lanl.gov/abs/hep-th/0009234}{{\tt
  hep-th/0009234}}].

\bibitem{Cardoso:2006xz}
G.~L. Cardoso, B.~de~Wit, and S.~Mahapatra, {\it {Black hole entropy functions
  and attractor equations}},  {\em JHEP} {\bf 03} (2007) 085,
  [\href{http://xxx.lanl.gov/abs/hep-th/0612225}{{\tt hep-th/0612225}}].

\bibitem{Almheiri:2014cka}
A.~Almheiri and J.~Polchinski, {\it {Models of AdS$_{2}$ backreaction and
  holography}},  {\em JHEP} {\bf 11} (2015) 014,
  [\href{http://xxx.lanl.gov/abs/1402.6334}{{\tt 1402.6334}}].

\bibitem{Maldacena:2016upp}
J.~Maldacena, D.~Stanford, and Z.~Yang, {\it {Conformal symmetry and its
  breaking in two dimensional Nearly Anti-de-Sitter space}},  {\em PTEP} {\bf
  2016} (2016), no.~12 12C104, [\href{http://xxx.lanl.gov/abs/1606.01857}{{\tt
  1606.01857}}].

\bibitem{Sarosi:2017ykf}
G.~S\'arosi, {\it {AdS$_{2}$ holography and the SYK model}},  {\em PoS} {\bf
  Modave2017} (2018) 001, [\href{http://xxx.lanl.gov/abs/1711.08482}{{\tt
  1711.08482}}].

\bibitem{Brown:2018bms}
A.~R. Brown, H.~Gharibyan, H.~W. Lin, L.~Susskind, L.~Thorlacius, and Y.~Zhao,
  {\it {Complexity of Jackiw-Teitelboim gravity}},  {\em Phys. Rev. D} {\bf 99}
  (2019), no.~4 046016, [\href{http://xxx.lanl.gov/abs/1810.08741}{{\tt
  1810.08741}}].

\bibitem{Cvetic:2016eiv}
M.~Cveti\v{c} and I.~Papadimitriou, {\it {AdS$_{2}$ holographic dictionary}},
  {\em JHEP} {\bf 12} (2016) 008,
  [\href{http://xxx.lanl.gov/abs/1608.07018}{{\tt 1608.07018}}]. [Erratum: JHEP
  01, 120 (2017)].

\bibitem{deWit:1984rvr}
B.~de~Wit, P.~G. Lauwers, and A.~Van~Proeyen, {\it {Lagrangians of N=2
  Supergravity - Matter Systems}},  {\em Nucl. Phys. B} {\bf 255} (1985)
  569--608.

\bibitem{deWit:1984wbb}
B.~de~Wit and A.~Van~Proeyen, {\it {Potentials and Symmetries of General Gauged
  N=2 Supergravity: Yang-Mills Models}},  {\em Nucl. Phys. B} {\bf 245} (1984)
  89--117.

\bibitem{Aniceto:2020saj}
P.~Aniceto, G.~L. Cardoso, and S.~Nampuri, {\it {$R^2$ corrected AdS$_2$
  holography}},  {\em JHEP} {\bf 03} (2021) 255,
  [\href{http://xxx.lanl.gov/abs/2010.08761}{{\tt 2010.08761}}].

\bibitem{Gibbons:1998fa}
G.~W. Gibbons and P.~K. Townsend, {\it {Black holes and Calogero models}},
  {\em Phys. Lett. B} {\bf 454} (1999) 187--192,
  [\href{http://xxx.lanl.gov/abs/hep-th/9812034}{{\tt hep-th/9812034}}].

\bibitem{Iyer:1994ys}
V.~Iyer and R.~M. Wald, {\it {Some properties of Noether charge and a proposal
  for dynamical black hole entropy}},  {\em Phys. Rev. D} {\bf 50} (1994)
  846--864, [\href{http://xxx.lanl.gov/abs/gr-qc/9403028}{{\tt
  gr-qc/9403028}}].

\bibitem{Papadimitriou:2010as}
I.~Papadimitriou, {\it {Holographic renormalization as a canonical
  transformation}},  {\em JHEP} {\bf 11} (2010) 014,
  [\href{http://xxx.lanl.gov/abs/1007.4592}{{\tt 1007.4592}}].

\bibitem{Skenderis:2002wp}
K.~Skenderis, {\it {Lecture notes on holographic renormalization}},  {\em
  Class. Quant. Grav.} {\bf 19} (2002) 5849--5876,
  [\href{http://xxx.lanl.gov/abs/hep-th/0209067}{{\tt hep-th/0209067}}].

\bibitem{Cardoso:2019wwf}
G.~L. Cardoso and T.~Mohaupt, {\it {Special geometry, Hessian structures and
  applications}},  {\em Phys. Rept.} {\bf 855} (2020) 1--141,
  [\href{http://xxx.lanl.gov/abs/1909.06240}{{\tt 1909.06240}}].

\end{thebibliography}
\end{document}